\documentclass[debug, journal]{rmaa_X}

\newcommand\rmxaa{RMAA}
\usepackage{paralist}

\usepackage{psfrag,color}

\usepackage[latin1]{inputenc}
\usepackage{afterpage,longtable,lscape}

\newcommand{\CS}[1]{\texttt{\textbackslash #1}}

\title{The San Pedro M\'artir Planetary Nebula \\ Kinematic Catalogue: Extragalactic Planetary Nebulae} 

\author{
  M. G. Richer,\altaffilmark{1} 
  J. A. L\'opez,\altaffilmark{1}
  E. D\'\i az-M\'endez,\altaffilmark{2}
  H. Riesgo,\altaffilmark{1}
  S.-H. B\'aez,\altaffilmark{3}
  Ma.-T. Garc\'\i a-D\'\i az,\altaffilmark{1}
  J. Meaburn\altaffilmark{4}
  D. M. Clark,\altaffilmark{1} 
  R. M. Calder\'on Olvera,\altaffilmark{5}
  G. L\'opez Soto,\altaffilmark{5}
  O. Toledano Rebolo\altaffilmark{6}
  }

\altaffiltext{1}{Instituto de Astronom\'\i a, UNAM, M\'exico.}

\altaffiltext{2}{Texas Christian University, Fort Worth, Texas, USA.}

\altaffiltext{3}{Facultad de F\'\i sica e Inteligencia Artificial, Universidad Veracruzana, M\'exico.}

\altaffiltext{4}{Jodrell Bank Observatory, University of Manchester, UK.}

\altaffiltext{5}{Universidad Nacional Autónoma de México, M\'exico.}

\altaffiltext{6}{Instituto de Ciencias F\'\i sicas, UNAM, M\'exico.}

\shortauthor{Richer et al.}
\shorttitle{SPM PN Kinematic Catalogue: Extragalactic Data}

\fulladdresses{
\item Sol-Haret B\'aez: Facultad de F\'\i sica e Inteligencia Artificial, Universidad Veracruzana, Circuito G. Aguirre Beltr\'an s/n, Zona Universitaria, Xalapa, Veracruz, M\'exico (solharet@hotmail.com).
\item Roxana Marisol Calder\'on Olvera, Giovanni L\'opez Soto:  Universidad Nacional Autónoma de México, Ciudad Universitaria, 04510 México, D.F., México  (roxanama@msn.com, blackhole\_mx@yahoo.com.mx).
\item David M. Clark, Mar\'\i a-Teresa Garc\'\i a-D\'\i az, Jos\'e Alberto L\'opez, Michael G. Richer, Hortensia Riesgo: OAN, Instituto de Astronom\'\i a, Universidad Nacional Aut\'onoma de M\'exico, P.O. Box 439027, San Diego, CA 92143 (\{tere, jal, richer, hriesgo\}@astrosmo.unam.mx).
\item Enrique D\'\i az-M\'endez: Dept. of Physics and Astronomy, Texas Christian University, Fort Worth, Texas, U.S.A.  76109 (e.d.mendez@tcu.edu).
\item John Meaburn: Jodrell Bank Observatory, Dept. of Physics \& Astronomy, University of Manchester, Macclesfield, Cheshire SK11 9DL, UK (jm@ast.man.ac.uk).
\item Oswaldo Toledano Rebolo: Instituto de Ciencias F\'\i sicas, Universidad Nacional Aut\'onoma de M\'exico, Av. Universidad s/n, Cuernavaca, Morelos, 62210, M\'exico (oswaldotoledano@hotmail.com).}
\listofauthors{M. G. Richer, J. A. L\'opez, E. D\'\i az-M\'endez, H. Riesgo, S.-H. B\'aez, Ma. T. Garc\'\i a-D\'\i az, J. Meaburn, D. M. Clark, R. M. Calder\'on Olvera, \& G. L\'opez Soto}
\indexauthor{Richer, M. G.}
\indexauthor{L\'opez, J. A.}
\indexauthor{D\'\i az-M\'endez, E.}
\indexauthor{Riesgo, H.}
\indexauthor{B\'aez, S.-H.}
\indexauthor{Garc\'\i a-D\'\i az, Ma. T.}
\indexauthor{Meaburn, J.}
\indexauthor{Clark, D. M.}
\indexauthor{Calder\'on Olvera, R. M.}
\indexauthor{L\'opez Soto, G.}
\indexauthor{Toledano Rebolo, O.}

\abstract{We present kinematic data for 211 bright planetary nebulae in eleven Local Group galaxies: M31 (137 PNe), M32 (13), M33 (33), Fornax (1), Sagittarius (3), NGC 147 (2), NGC 185 (5), NGC 205 (9), NGC 6822 (5), Leo A (1), and Sextans A (1).  The data were acquired at the Observatorio Astron\'omico Nacional in the Sierra de San Pedro M\'artir using the 2.1m telescope and the Manchester Echelle Spectrometer in the light of [\ion{O}{3}]$\lambda$5007 at a resolution of 11\,km\,s$^{-1}$.  A few objects were observed in H$\alpha$.  The internal kinematics of bright planetary nebulae do not depend strongly upon the metallicity or age of their progenitor stellar populations, though small systematic differences exist.  The nebular  kinematics and H$\beta$ luminosity require that the nebular shells be accelerated during the early evolution of their central stars.  Thus, kinematics provides an additional argument favoring similar stellar progenitors for bright planetary nebulae in all galaxies.
}

\resumen{Presentamos datos cinem\'aticos para 211 nebulosas planetarias brillantes en once galaxias del Grupo Local: M31 (137 NPs), M32 (13), M33 (33), Fornax (1), Sagittarius (3), NGC 147 (2), NGC 185 (5), NGC 205 (9), NGC 6822 (6), Leo A (1), y Sextans A (1).  Adquirimos los datos en el Observatorio Astron\'omico Nacional en la Sierra de San Pedro M\'artir con el telescopio de 2.1m y el Espectr\'ometro Manchester Echelle en la l\'\i nea de [\ion{O}{3}]$\lambda$5007 con una resoluci\'on de 11\,km\,s$^{-1}$. Observamos algunos objetos en H$\alpha$.  La cinem\'atica de nebulosas planetarias brillantes no depende fuertemente de la metalicidad o la edad de la poblaci\'on estelar progenitora en sus galaxias hu\'espedes, aunque existen peque\~nas diferencias sistem\'aticas.  La cinem\'atica y la luminosidad en H$\beta$ de las c\'ascaras nebulares requieren la aceleraci\'on de las c\'ascaras durante la evoluci\'on temprana de las estrellas centrales.  As\'\i, la cinem\'atica representa otro argumento en favor de estrellas progenitoras similares para las nebulosas planeatarias brillantes en todas galaxias.
}

\addkeyword{ISM: kinematics and dynamics}
\addkeyword{Local Group}
\addkeyword{Planetary Nebulae}
\addkeyword{Stars: Evolution}

\begin{document}
\maketitle

\section{Introduction}\label{introduction}

Planetary nebulae are the immediate descendants of asymptotic giant branch (AGB) stars of low and intermediate masses ($1\, M_{\odot} < M< 8\,M_{\odot}$).  The mass lost on the AGB (or part of it) is seen as the ionized nebular shell in planetary nebulae.  The composition of these nebular shells is extremely useful in studying the nucleosynthetic production of their precursor stars.  The cycling of matter through these stars and its transformation is part of the chemical evolution of galaxies, since the progenitors of planetary nebulae are responsible for much of the helium, carbon, nitrogen, and some s-process elements in the universe.  

The luminosity function of bright extragalactic planetary nebulae (PNLF) has been used extensively as a distance indicator for about two decades \citep{jacoby1989, ciardulloetal1989}.  While some progress has been made on understanding the nature of the progenitor stars of bright planetary nebulae \citep[e.g.,][]{richermccall2008}, the comments made by \citet{pottasch1990} are still largely valid: \lq\lq It may seem rather strange to determine the distance to the galactic centre by calibrating against the much more distant galaxy M31, but it is no stranger than the idea of using PN as standard candles in the first place, since individual distances to PN are so poorly known."  

Bright extragalactic planetary nebulae have two key advantages for evolutionary studies with respect to their Galactic counterparts.  First, their distances are known, so their absolute luminosities are studied easily.  Second, in galaxies more distant than the Magellanic Clouds, planetary nebulae are unresolved for ground-based observations, making it easy to study their integrated spectral properties, even at high spectral resolution.  The drawback, of course, is the lack (usually) of spatial resolution available for Galactic planetary nebulae that is so useful in studying physical processes.  

Over the past decade, a substantial quantity of low resolution spectroscopy has been acquired \citep[e.g.,][]{jacobyciardullo1999, walshetal1999, rothetal2004, mendezetal2005, penaetal2007, richermccall2008, magrinigoncalves2009}.  On the other hand, the only high resolution spectroscopy of extragalactic planetary nebulae suitable for studying their internal kinematics is that of \citet{zijlstraetal2006} and \citet{arnaboldietal2008}, apart from the pioneering efforts of \citet{dopitaetal1985, dopitaetal1988}.  Here, we present our high resolution spectroscopic observations of 211 planetary nebulae in 11 Local Group galaxies.  For the first time, these data allow redundant comparisons across different stellar populations.  This study is part of a larger effort to understand the systematics of planetary nebula kinematics within our Milky Way and the Local Group.  \citet{lopezetal2010} present our results for Galactic planetary nebulae.  

We present our observations and the data reduction in \S \ref{observations}.  We explain the analysis as well as its limitations in \S \ref{analysis}.  The results follow in \S\ref{results}.  We argue that the [\ion{O}{3}]$\lambda 5007$ line width is an adequate description of the kinematics of the majority of the ionized gas.  We demonstrate that our radial velocities are accurate and in agreement with extant data.  We find that the average line widths for bright planetary nebulae in all galaxies are similar, though there is a trend of decreasing line width with nuclear distance in the disc of M31.  In \S \ref{discussion}, we consider the implications of the foregoing, with the most important being that the progenitor stars of bright planetary nebulae in all galaxies span a relatively small range in mass and that the central star plays an important role in accelerating the ionized shells of these objects.  \S \ref{conclusions} summarizes our conclusions.

Here, we focus on what we can learn of the evolution of bright extragalactic planetary nebulae and their progenitor stars from their [\ion{O}{3}]$\lambda 5007$ line widths.  We make no attempt to investigate the kinematics of individual objects nor to interpret the line widths in terms of internal kinematics, as our spatially-integrated line profiles make this very difficult.  

\section{Observations and Data Reduction}\label{observations}

The observations were obtained during twelve observing runs between 2001 September and 2007 August (see Table \ref{tab_obs_run}).  All of the data were acquired with the 2.1m telescope  at the Observatorio Astronómico Nacional in the Sierra de San Pedro Mártir (OAN-SPM) and the Manchester Echelle Spectrometer \citep[MES-SPM; ][]{meaburnetal1984, meaburnetal2003}.  The MES-SPM is a long slit echelle spectrometer, but uses narrow-band filters, instead of a cross-disperser, to isolate the orders containing the emission lines of interest.  In our case, filters isolated orders 87 and 114 containing the H$\alpha$ and [\ion{O}{3}]$\lambda 5007$ emission lines, respectively.  All observations used a 150\,$\mu$m wide slit, equivalent to $1\farcs9$ on the sky.  When coupled with a SITe $1024\times 1024$ CCD with 24\,$\mu$m pixels binned $2\times 2$, the resulting spectral resolutions were approximately 0.077\,\AA/pix and 0.100\,\AA/pix at [\ion{O}{3}]$\lambda 5007$ and H$\alpha$, respectively (equivalent to 11\,km/s for 2.6\,pix FWHM).  Immediately before or after every object spectrum, exposures of a ThAr lamp were taken to calibrate in wavelength.  The internal precision of the arc lamp calibrations is better than $\pm 1.0$\,km/s.

\begin{table}[!t]\centering
  \setlength{\tabnotewidth}{0.5\columnwidth}
  \tablecols{2}
  \caption{Observing Runs} \label{tab_obs_run}
 \begin{tabular}{ll}
    \toprule
    Run & Dates \\
    \midrule
2001sep & 2001 Sep 21-26 \\
2002jul & 2002 Jul 23-24 \\
2002oct & 2002 Oct 31-2002 Nov 05 \\
2003oct & 2003 Oct 14-19 \\
2004jun & 2004 Jun 10-16 and 20-26 \\
2004nov & 2004 Nov 19-25 \\
2004dec & 2004 Dec 17-18 \\
2005jul & 2005 Jul 22-31 \\
2005sep & 2005 Sep 10-15 \\
2006sep & 2006 Sep 05-12 \\
2007feb & 2007 Feb 01-08 \\
2007aug & 2007 Aug 18-27 \\
    \bottomrule
  \end{tabular}
\end{table}

All spectra were of 30 minutes duration.  Depending upon the resulting signal-to-noise, up to four spectra were acquired, occasionally during more than one observing run.  With very few exceptions (PN24 and PN25 in M32), the spectrometer slit was always oriented north-south over the object(s) of interest.  

Table \ref{tab_o3} presents our entire sample of extragalactic planetary nebulae.  In total, we observed 211 extragalactic planetary nebulae in 11 Local Group galaxies.  More than half of the sample is drawn from M31 (137 objects), with M33 providing the second largest sample (33 objects).  The remaining galactic hosts are dwarfs and their planetary nebula populations are smaller: one planetary nebula was observed in each of Fornax, Leo A, and Sextans A, 2 in NGC 147, 3 in Sagittarius, 5 in NGC 185, 6 in NGC 6822, 9 in NGC 205, and 13 in M32.  Most of the objects we observed are within two magnitudes of the peak of the PNLF, though we did try to observe fainter objects if low-resolution spectra existed in the literature.  Table \ref{tab_Ha} presents our few H$\alpha$ spectra.  In Table \ref{tab_cHII}, we present observations of three objects whose classifications as planetary nebulae are dubious.  

The object names we adopt in Tables \ref{tab_o3}-\ref{tab_cHII} come from the following sources: Leo A and Sextans A from \citet{jacobylesser1981}; Fornax from \citet{danzigeretal1978}; M31 from \citet{fordjacoby1978}, \citet{lawrieford1982}, \citet{noltheniusford1987}, \citet{ciardulloetal1989}, \citet{richeretal2004}, and \citet{merrettetal2006}; M32 from \citet{fordjenner1975} and \citet{ford1983}; M33 from \citet{magrinietal2000}; NGC 147 from \citet{corradietal2005}; NGC 185 from \citet{fordetal1977}; NGC 205 from \citet{fordetal1973}, \citet{ford1978}, and \citet{ciardulloetal1989}; NGC 3109 from \citet{penaetal2007}; NGC 6822 from \citet{killendufour1982} and \citet{leisyetal2005}; and Sagittarius from \citet{ackeretal1992}.  The names follow strict historical precedent, except for NGC 147, NGC 3109 \citep[we do not use][]{richermccall1992}, and perhaps some objects in M31 from \citet{noltheniusford1987} whose identifications are taken from \citet{merrettetal2006}.  Coordinates for the objects from \citet{richeretal2004} and a new planetary nebula candidate that we discovered in M33 are given in Table \ref{tab_RLH}.  

\begin{table}[!t]\centering
  \setlength{\tabnotewidth}{0.9\columnwidth}
  \tablecols{4}
  \caption{New Planetary nebulae in M31 and M33} \label{tab_RLH}
 \begin{tabular}{llcc}
    \toprule
Galaxy  &  Object & $\alpha (2000)$ & $\delta (2000)$ \\
    \midrule
M31 & f08n04 & 00 49 50 & 42 31 40 \\
M31 & f08n05 & 00 49 32 & 42 37 38 \\
M31 & f08n08 & 00 50 13 & 42 29 48 \\
M31 & f08n09 & 00 48 48 & 42 51 36 \\
M31 & f08n10 & 00 47 54 & 42 14 58 \\
M31 & f15n12002 & 00 47 46.8 & 42 12 15 \\
M31 & f29n2065 & 00 33 41.1 & 39 31 51 \\
M31 & f29n9178 & 00 37 21.1 & 39 50 51 \\
M31 & f32n1 & 00 49 55 & 38 32 49 \\
M33 & NewPN & 01 32 56 & 30 25 56 \\
    \bottomrule
        \tabnotetext{}{The objects in M31 are from \citet{richeretal2004}.}
  \end{tabular}
\end{table}

All of the spectra were reduced using the specred package of the Image Reduction and Analysis Facility\footnote{IRAF is distributed by the National Optical Astronomical Observatories, which is operated by the Associated Universities for Research in Astronomy, Inc., under contract to the National Science Foundation.} (IRAF) following \citet{masseyetal1992}.  We edited each spectrum for the presence of cosmic rays.  We then extracted the source spectra and used these apertures to extract ThAr spectra from the lamp spectra.  The latter were used to calibrate in wavelength.  If more than one object spectrum was obtained, they were co-added after being calibrated in wavelength.  If they did not coincide exactly in wavelength, likely the result of centering differently or low signal-to-noise (S/N), they were shifted to a common wavelength and then co-added.  Note that shifting the spectra before co-adding minimizes the eventual line width that we measure.  We did not calibrate in flux.  

\begin{figure}[!t]
\begin{center}
  \includegraphics[width=0.30\columnwidth]{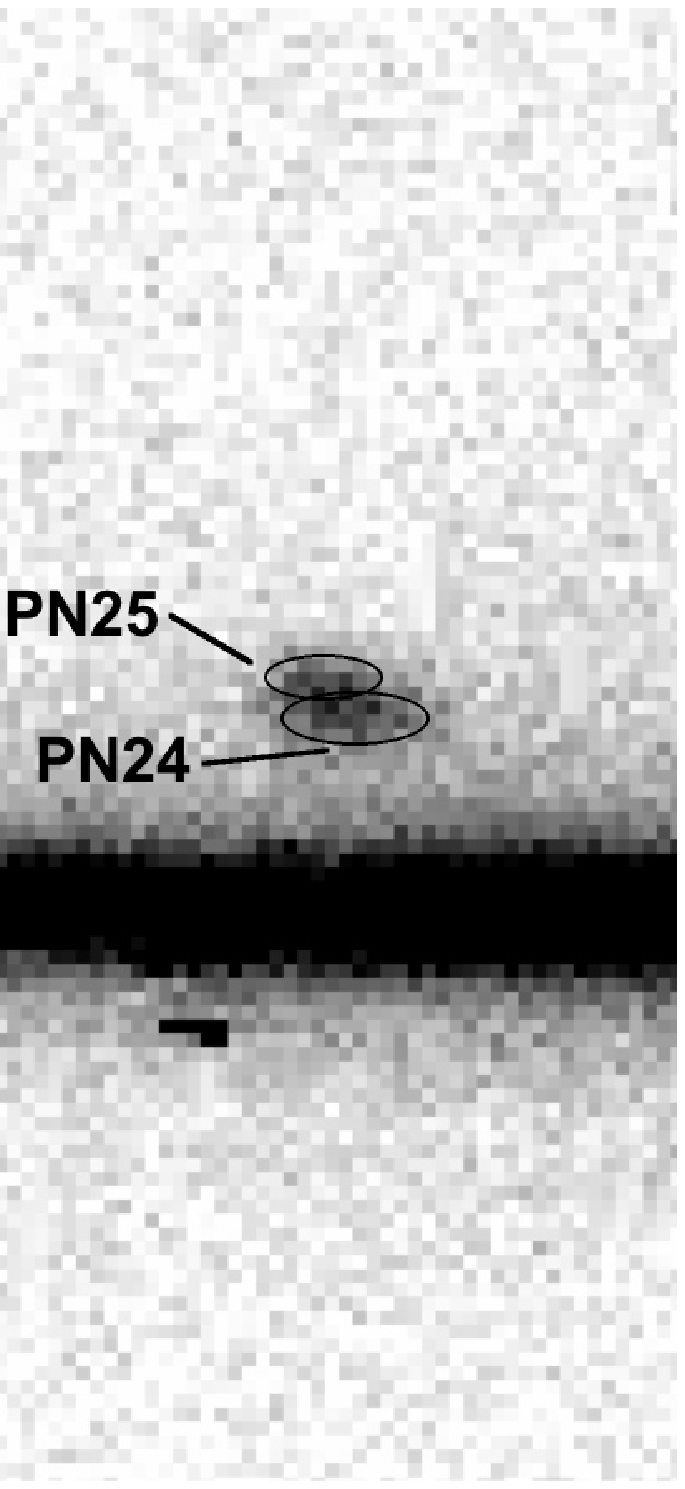}
  \includegraphics[width=0.30\columnwidth]{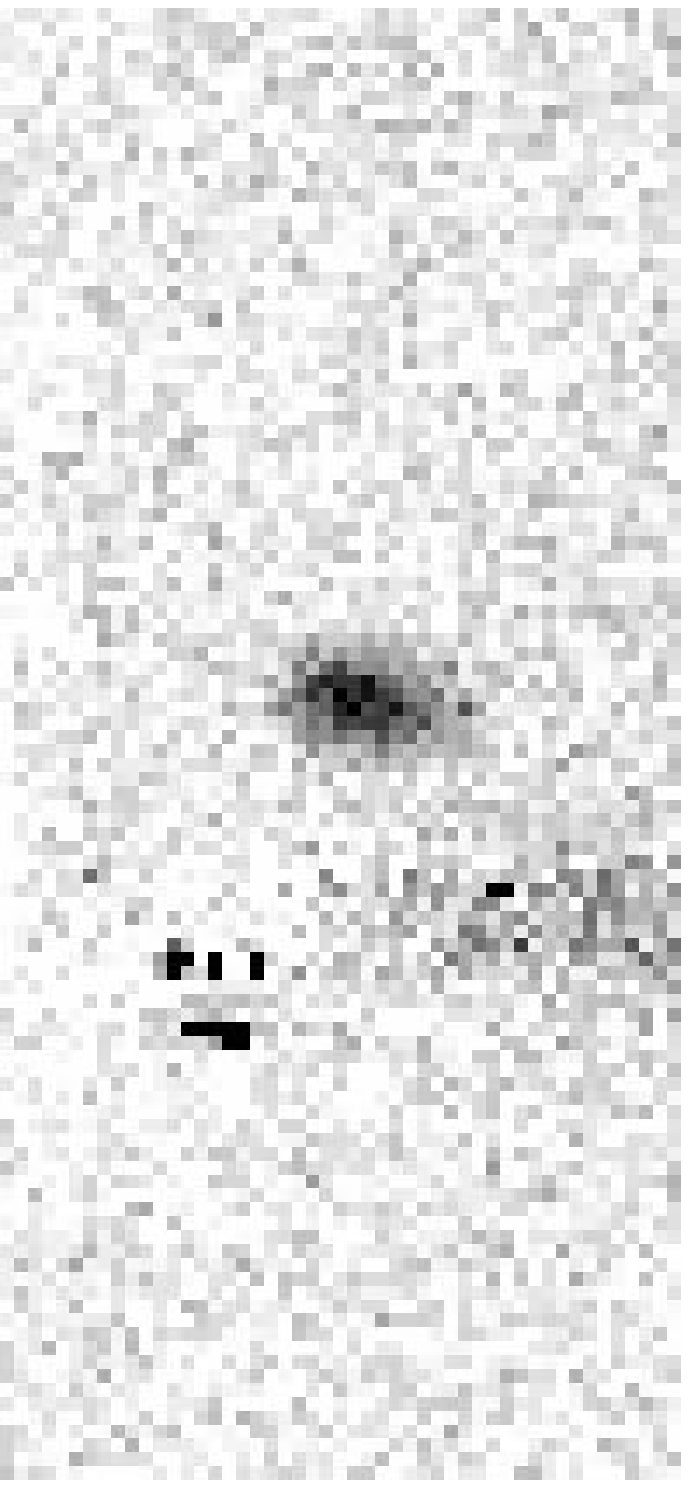}
  \includegraphics[width=0.33\columnwidth]{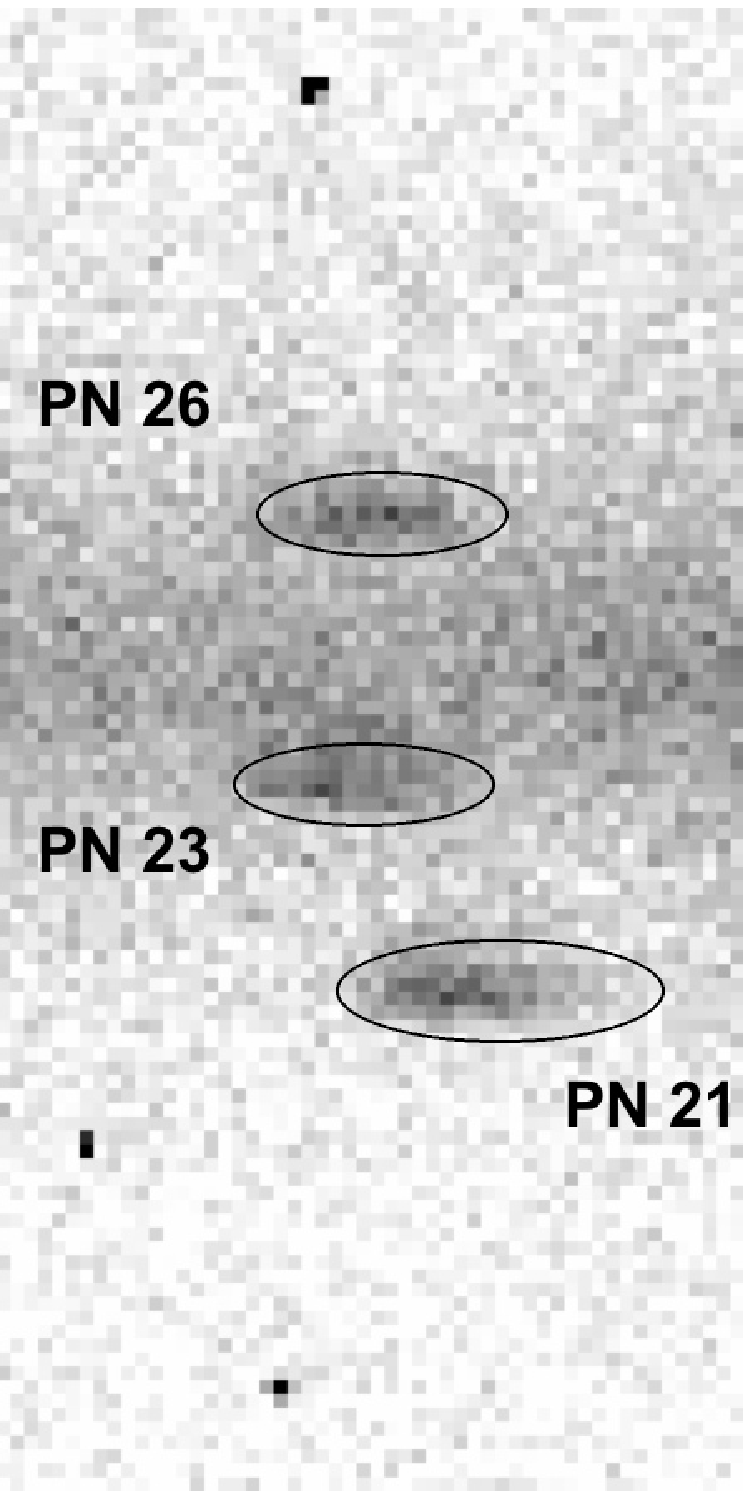}
\end{center}
  \caption{The left panel presents the raw two-dimensional spectrum for PN24 and PN25 in M32 (east is up, bluer wavelengths to the left).  The continuous emission below them (to the west) is from M32's nuclear region.  The planetary are sufficiently close together that their emission is merged into a single teardrop-shaped profile.  The middle panel presents the same spectrum for PN24 and PN25 once M32's continuous emission had been filtered out.  For comparison, the right panel presents the two-dimensional spectrum for PN26, PN23, and PN21 in M32 (ordered north to south; north is up, bluer wavelengths to the left) at the same spatial scale and taken immediately prior to the spectrum of PN24 and PN25 (continuous emission from M32 is again visible).  The two-dimensional (spatially-merged) spectrum of PN24 and PN25 is clearly more complex than the spectra of PN26, PN23, or PN21.  PN25 is to the east (above) of PN24, has a more blueshifted radial velocity, and a narrower line profile (see Table \ref{tab_o3}), the two effects explaining the tilted teardrop shape of the combined profile.  The \lq\lq individual" profiles for PN24 and PN25 were obtained by splitting the combined emission at the spatial center.}  
\label{fig_m32pn24pn25}
\end{figure}

Except for PN24 and PN25 in M32, all of the objects were observed as spatially distinct sources.  However, there were occasions when multiple objects fell within the spectrometer slit, particularly for M31 and M32 (see Fig. \ref{fig_m32pn24pn25}, right panel).  PN24 and PN25 in M32 required special treatment, since their images were not clearly resolved spatially by the spectrometer (the spectrometer slit was oriented east-west in the hope of better separating them).  Fortunately, both their radial velocities and line widths are distinct, which helps separating their spectra.  Fig. \ref{fig_m32pn24pn25} presents the two-dimensional spectrum of PN24 and PN25 (both raw and filtered so as to remove M32's continuous spectrum) and compares it with the two-dimensional spectrum of PN26, PN23, and PN21 (in M32) taken immediately before.  It is clear that the two-dimensional spectrum of PN24 and PN25 is more complicated in the spatial direction (vertical in Fig. \ref{fig_m32pn24pn25}) than those of the other PNe.  The \lq\lq individual" profiles for PN24 and PN25 were obtained by splitting the combined profile at the spatial centre since the two PNe are similarly bright \citep{ciardulloetal1989}.  Table \ref{tab_o3} presents the analysis of both the combined and split profiles for PN24 and PN25.  

\section{Data Analysis}\label{analysis}

\begin{figure*}[!t]
\begin{center}
  \includegraphics[width=\columnwidth]{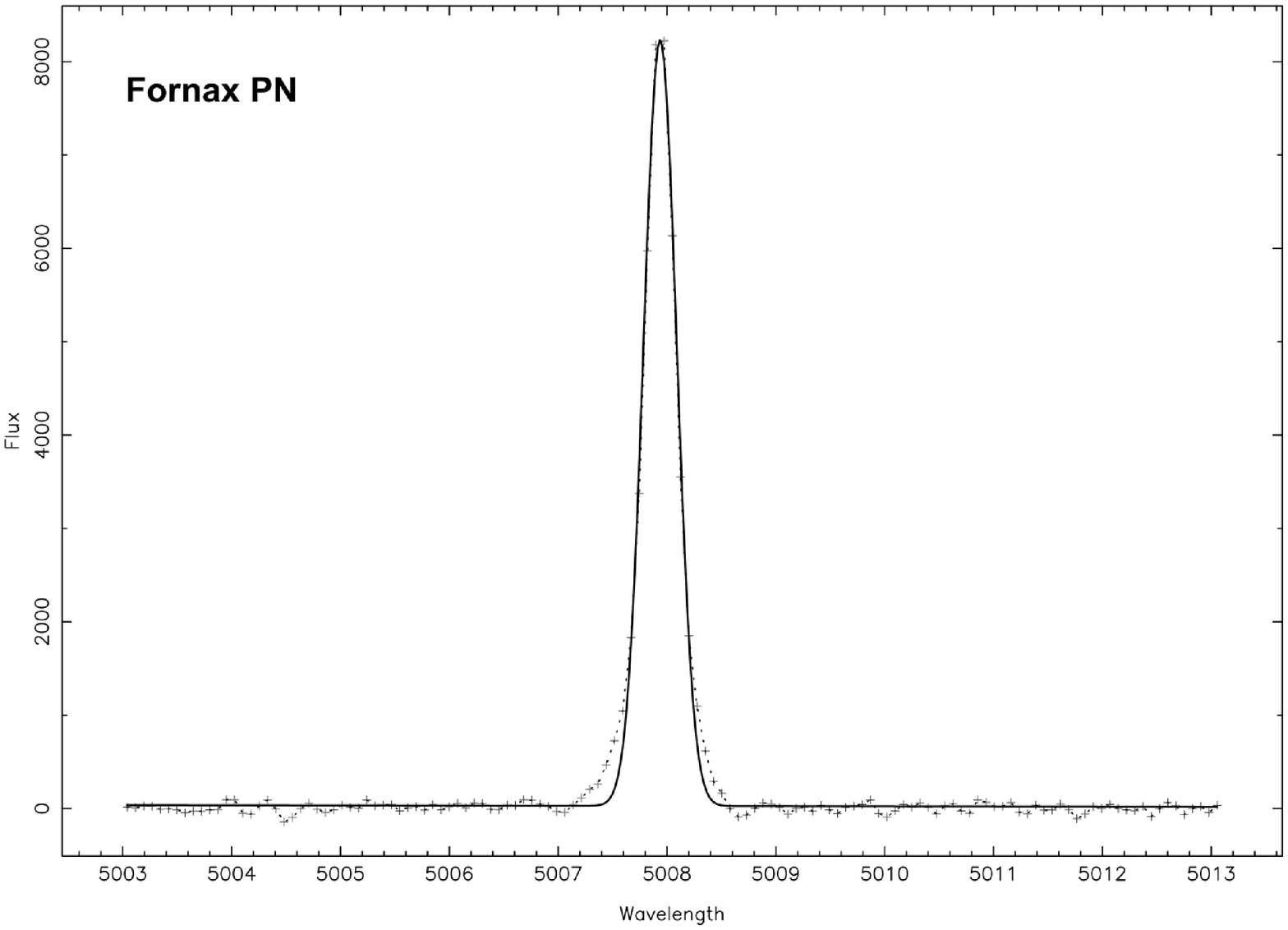}
  \includegraphics[width=\columnwidth]{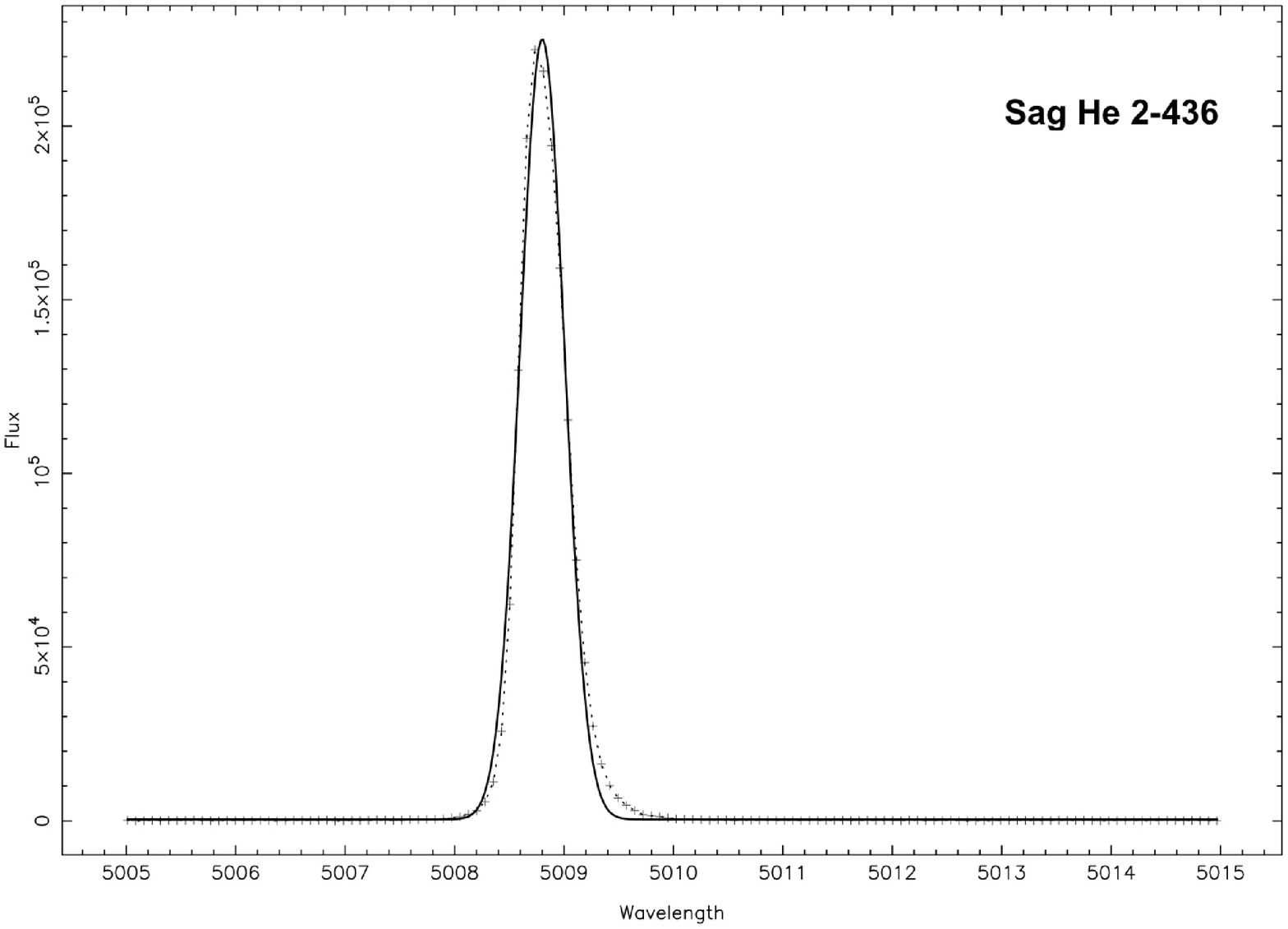}
  \includegraphics[width=\columnwidth]{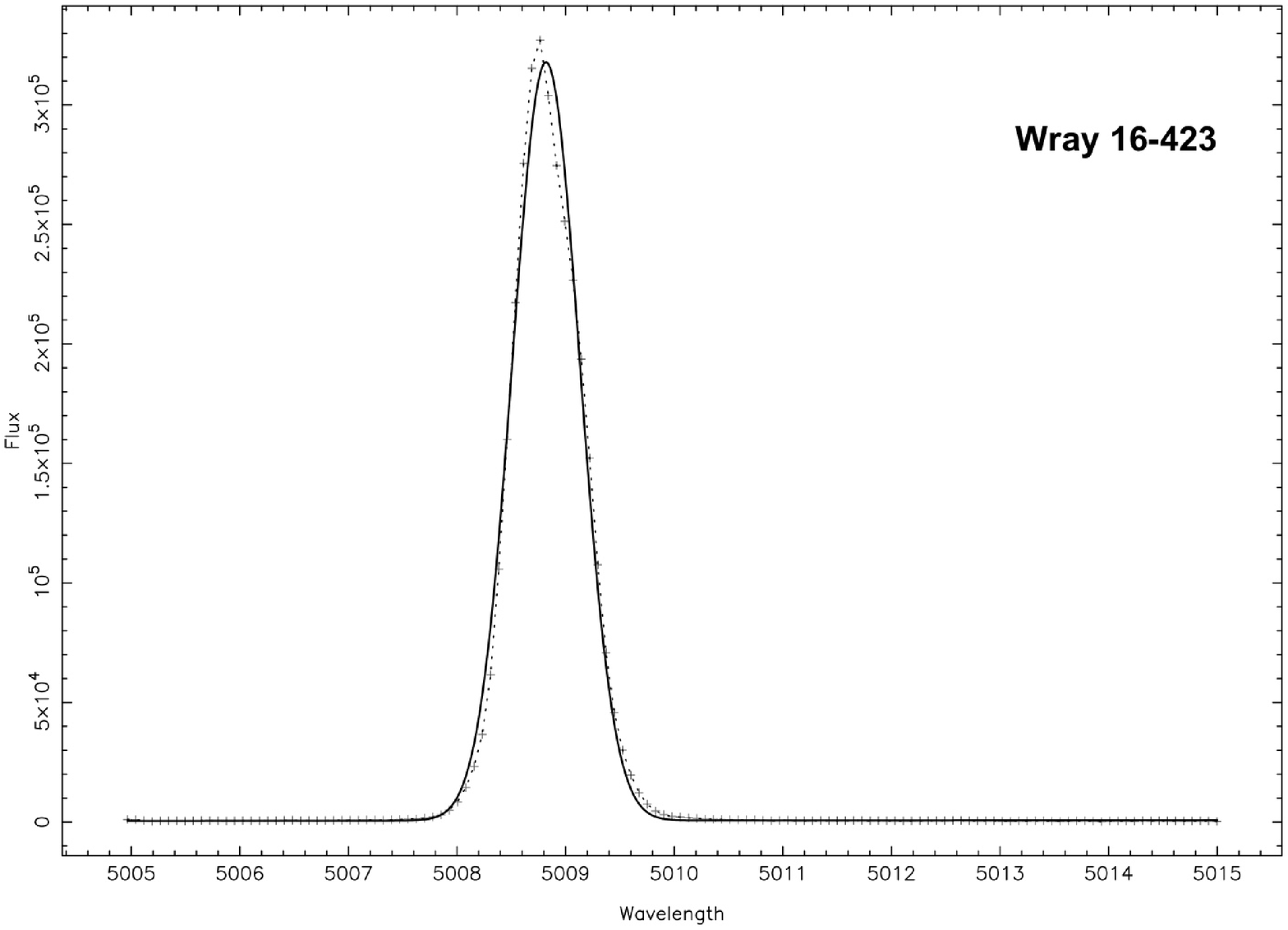}
  \includegraphics[width=\columnwidth]{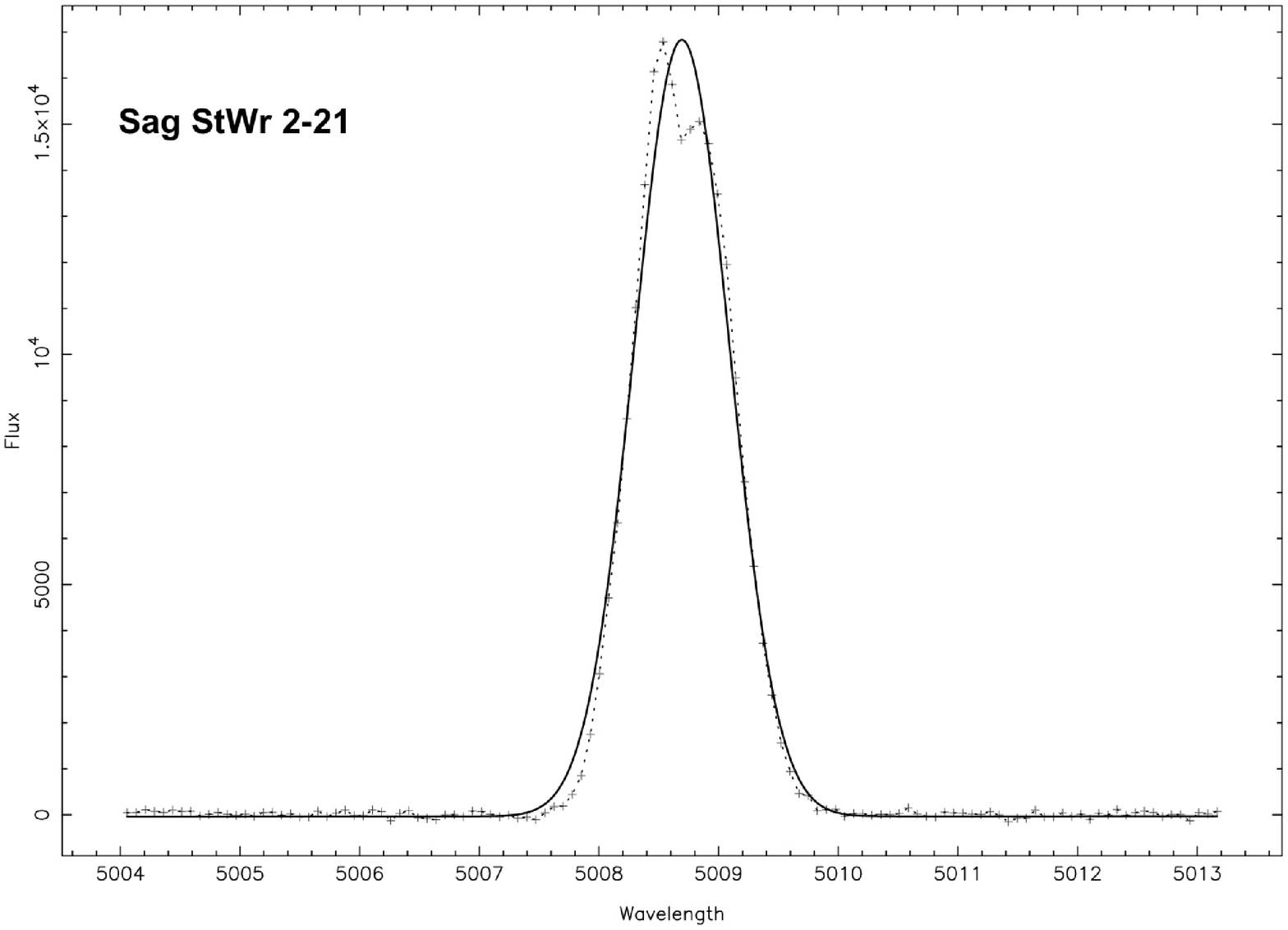}
  \includegraphics[width=\columnwidth]{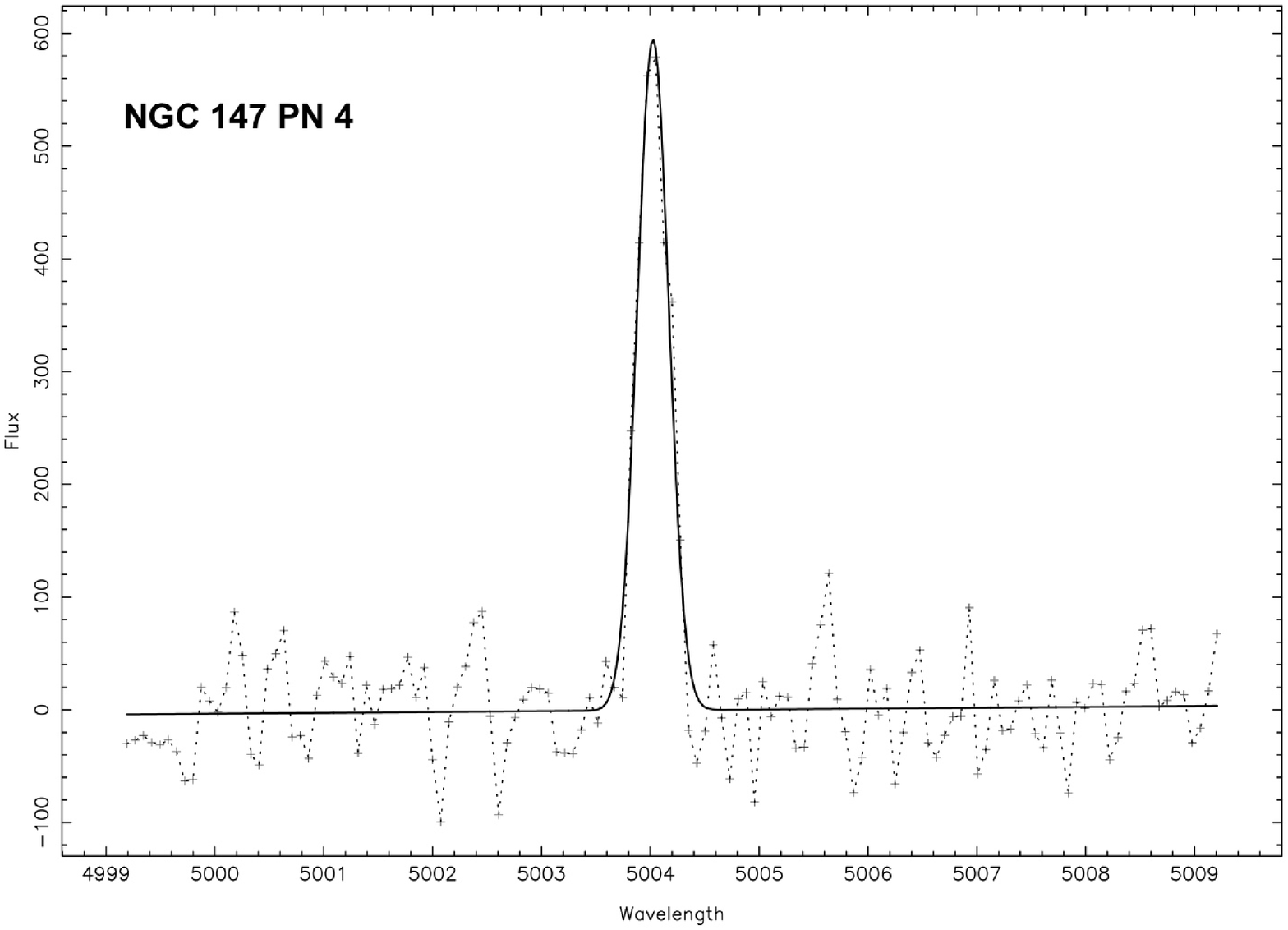} 
  \includegraphics[width=\columnwidth]{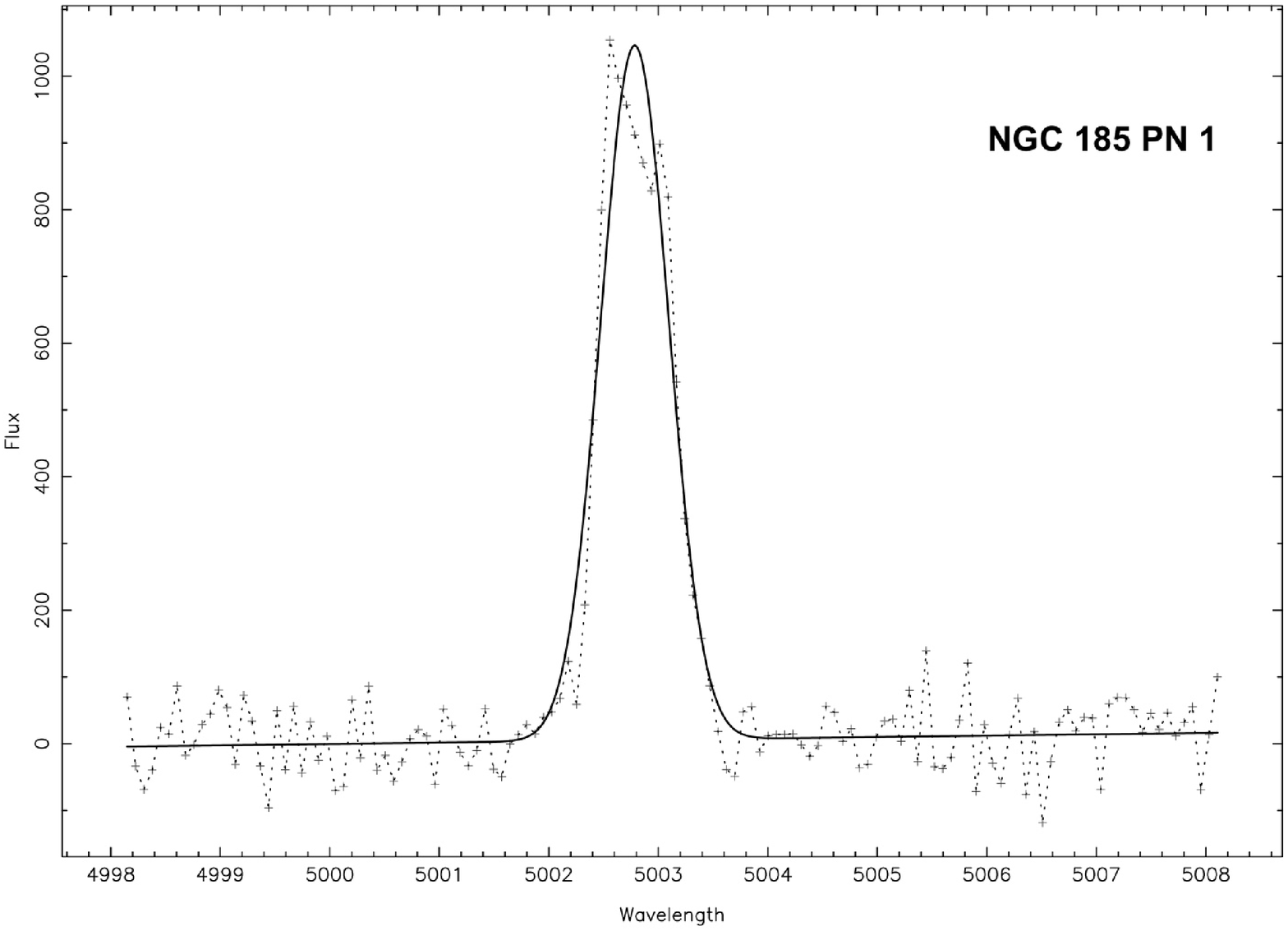}
\end{center}
  \caption{This and the following two figures present a gallery of representative line profiles for extragalactic planetary nebulae.  Here, the line profiles for planetary nebulae in four dwarf spheroidals are shown.  The light-colored symbols are the data and the solid line is the Gaussian fit.  In all cases, a 10\AA\ spectral interval is plotted.  The line profiles for the planetary nebulae in Fornax and Sagittarius are those with the highest S/N.  Even so, deviations from a Gaussian profile are small, though real (8-11\% of the total flux).  For the remaining objects, the deviations are usually insignificant.}
\label{fig_line_prof_gallery1}
\end{figure*}

\begin{figure*}[!t]
\begin{center}
  \includegraphics[width=\columnwidth]{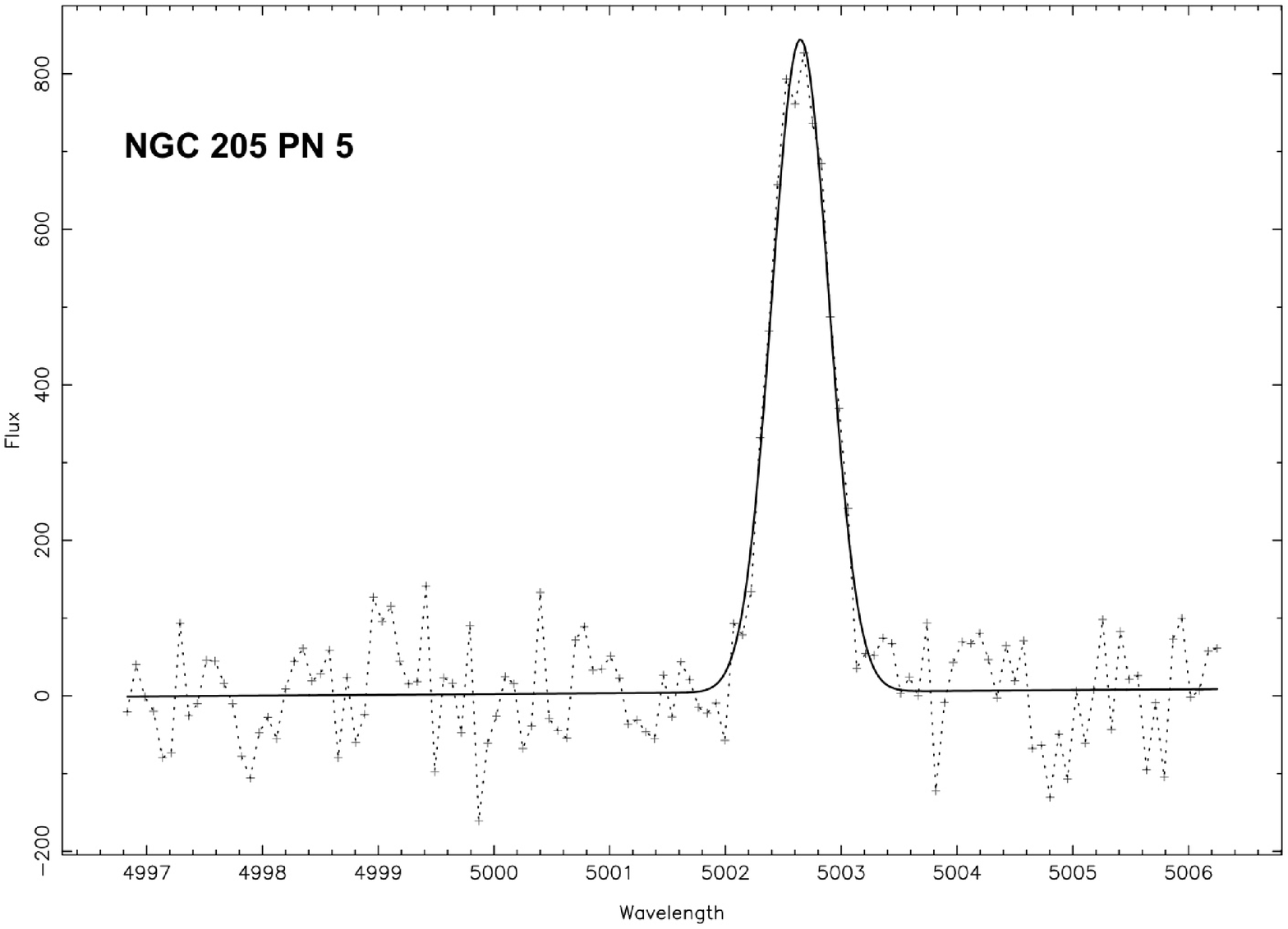}
  \includegraphics[width=\columnwidth]{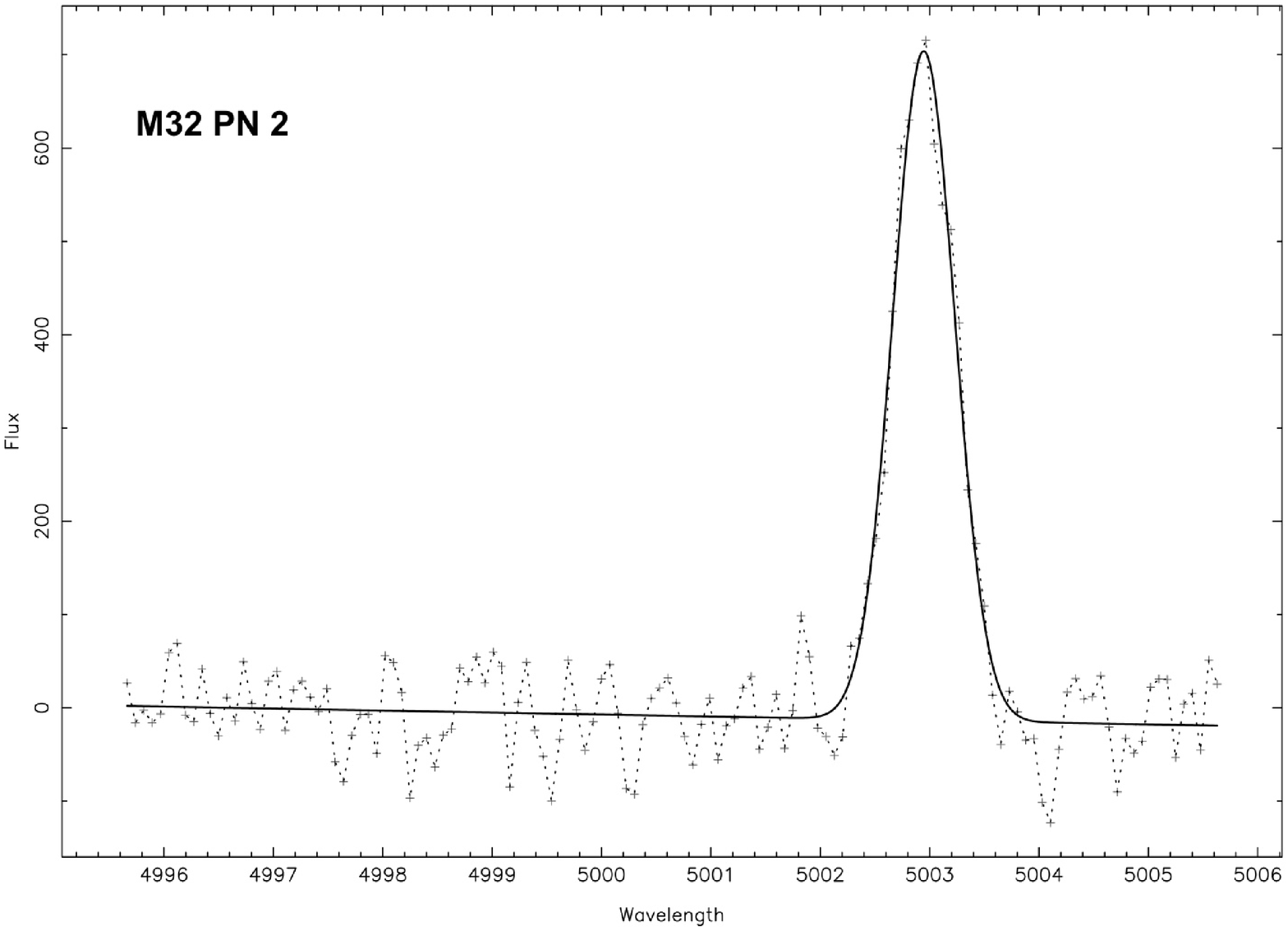}
  \includegraphics[width=\columnwidth]{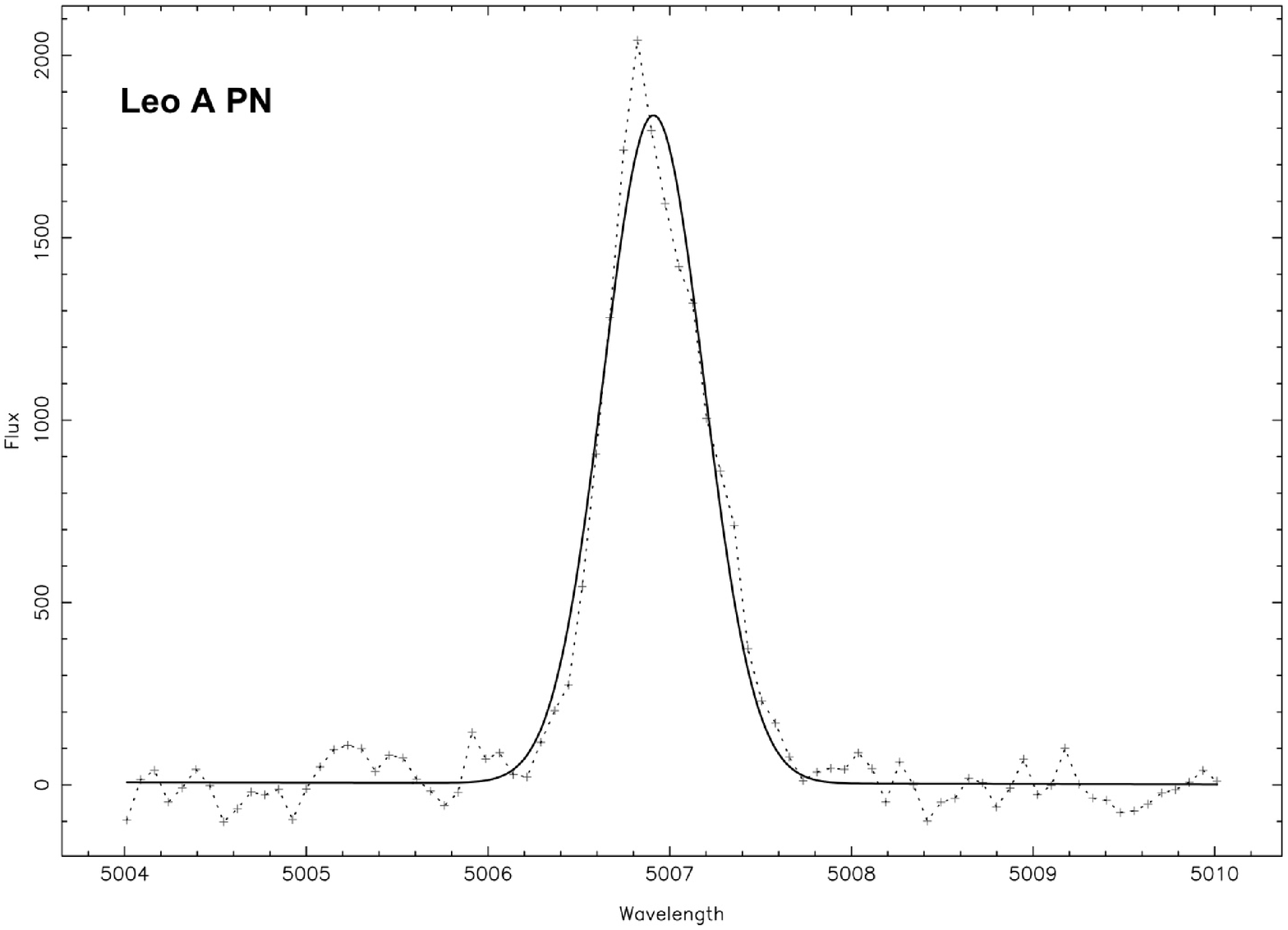} 
  \includegraphics[width=\columnwidth]{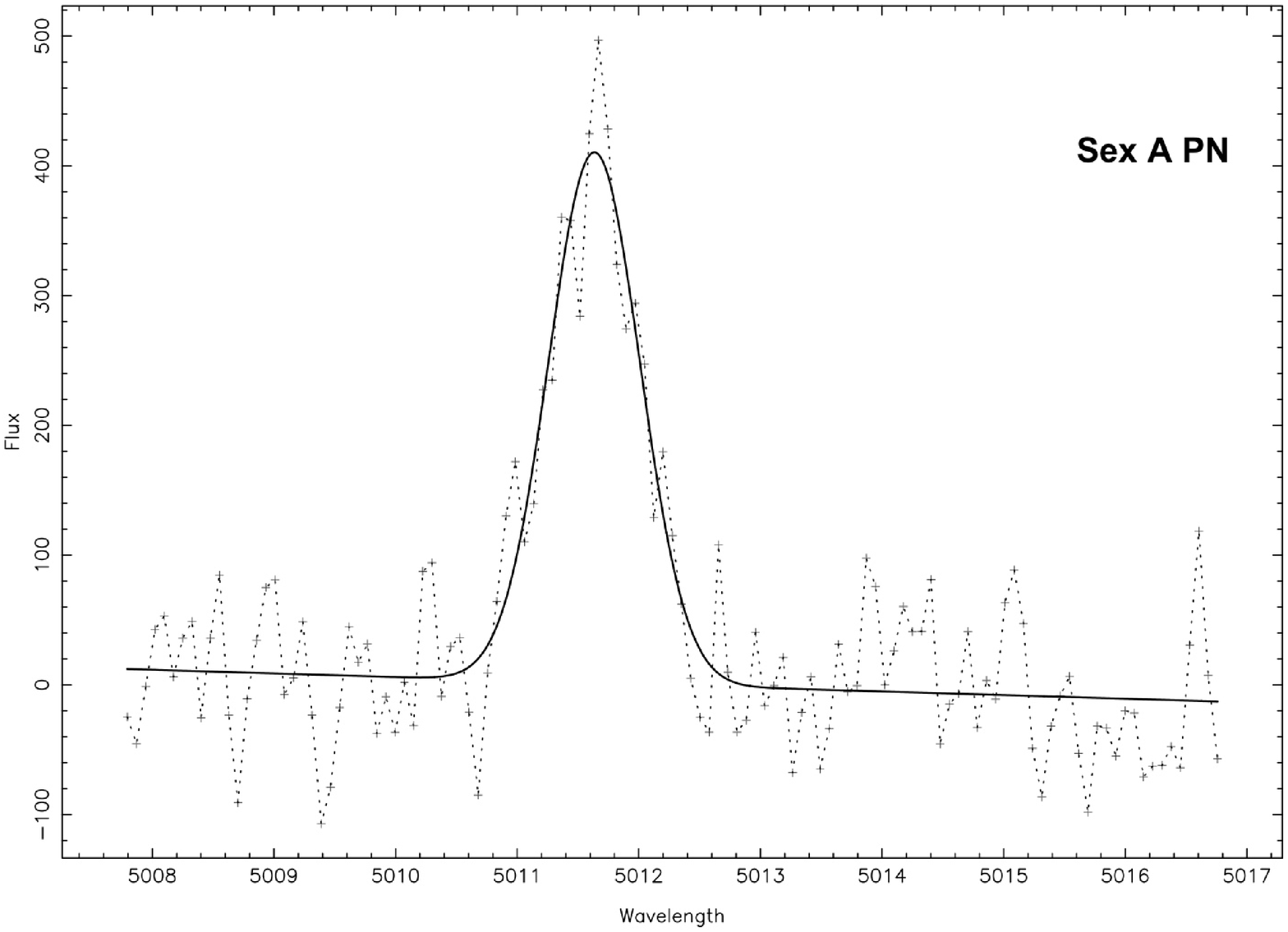}
  \includegraphics[width=\columnwidth]{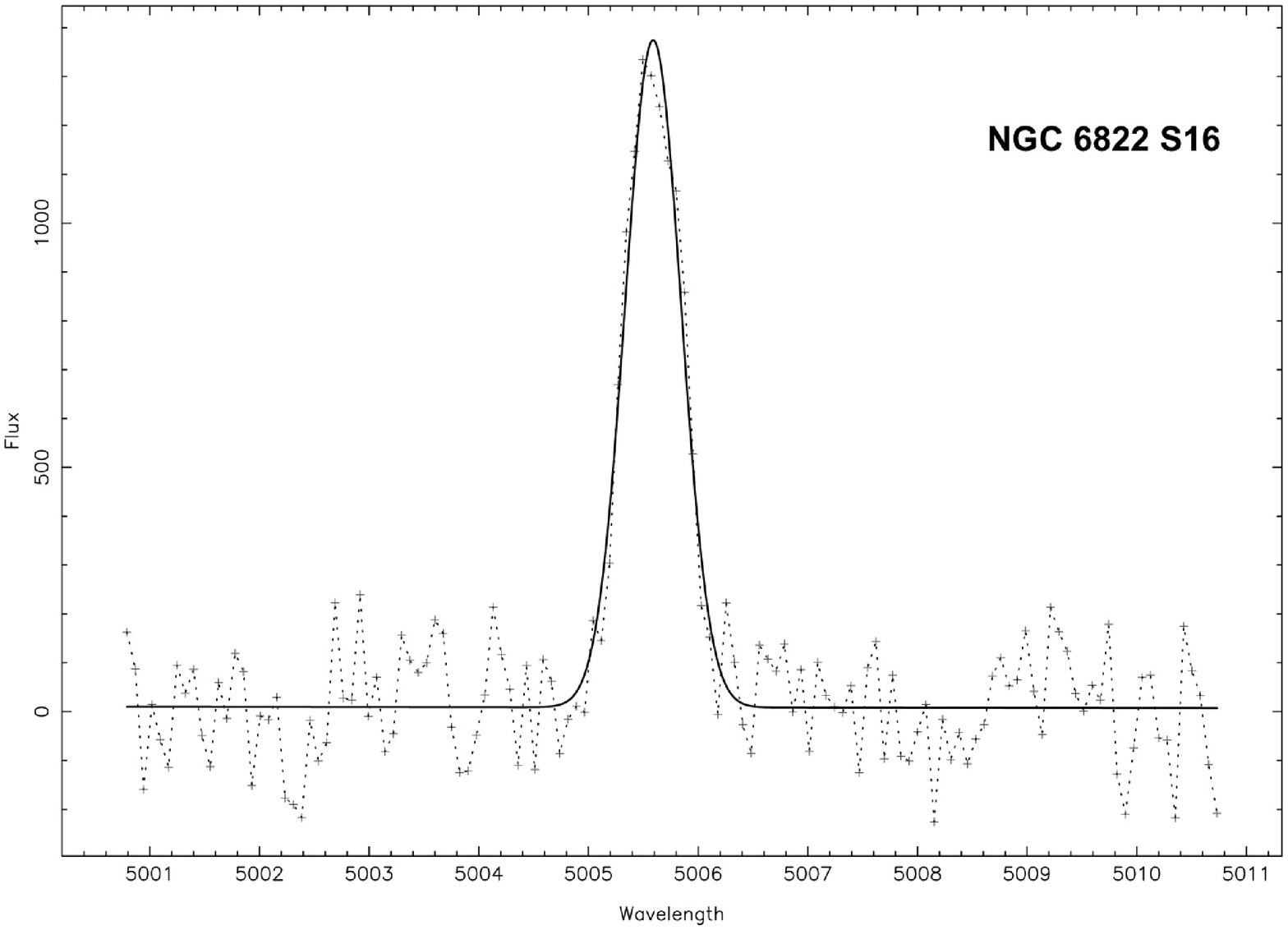}
  \includegraphics[width=\columnwidth]{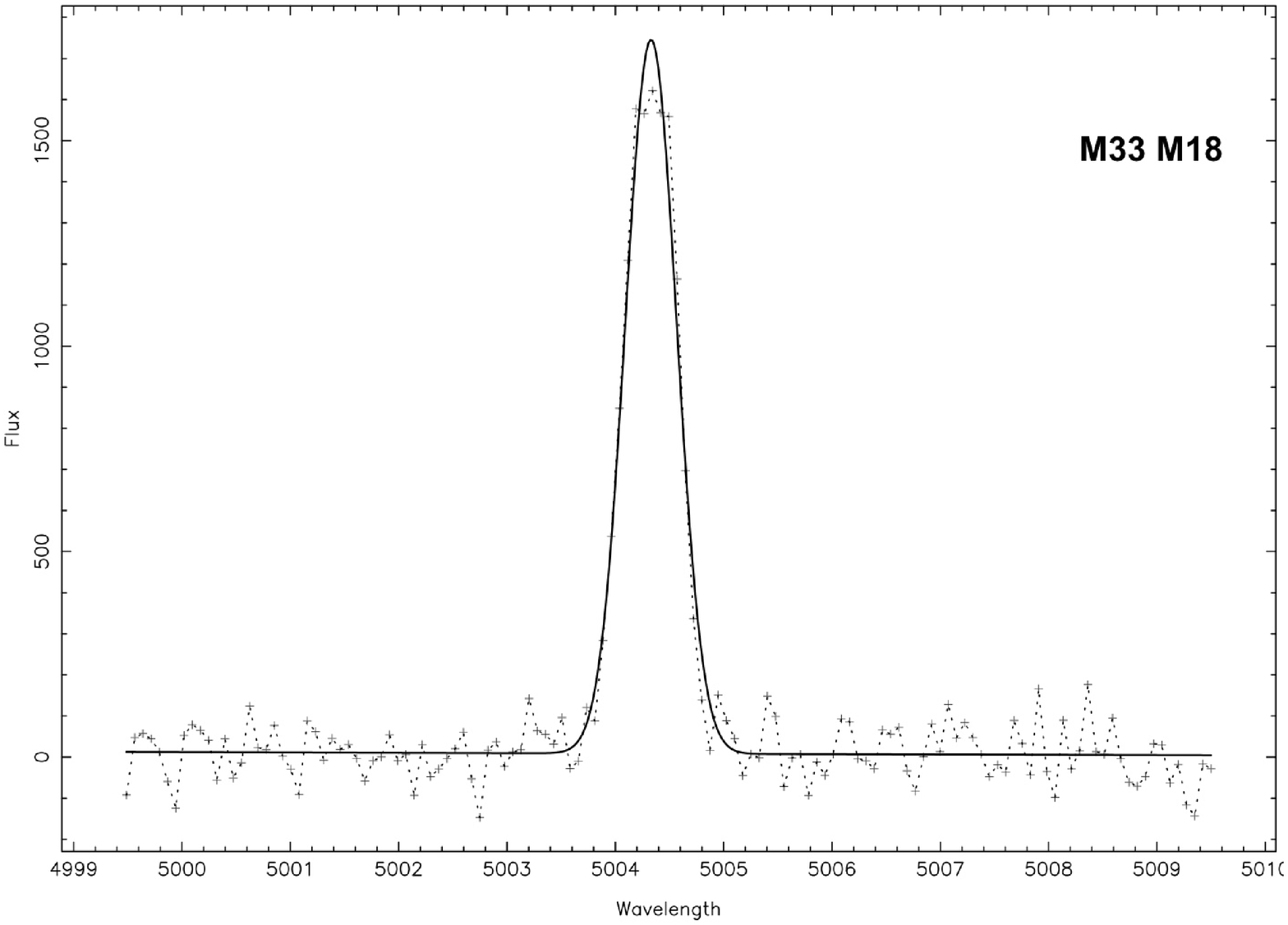}
\end{center}
  \caption{The line profiles are shown for planetary nebulae in NGC 205, M32, the dwarf irregular galaxies Leo A, Sex A, and NGC 6822, and the spiral galaxy M33.  Deviations of the data about the Gaussian fit are insignificant, even though these are among the profiles with the best S/N in these galaxies.  See Fig. \ref{fig_line_prof_gallery1} for further details.}
\label{fig_line_prof_gallery2}
\end{figure*}

\begin{figure*}[!t]
\begin{center}
  \includegraphics[width=\columnwidth]{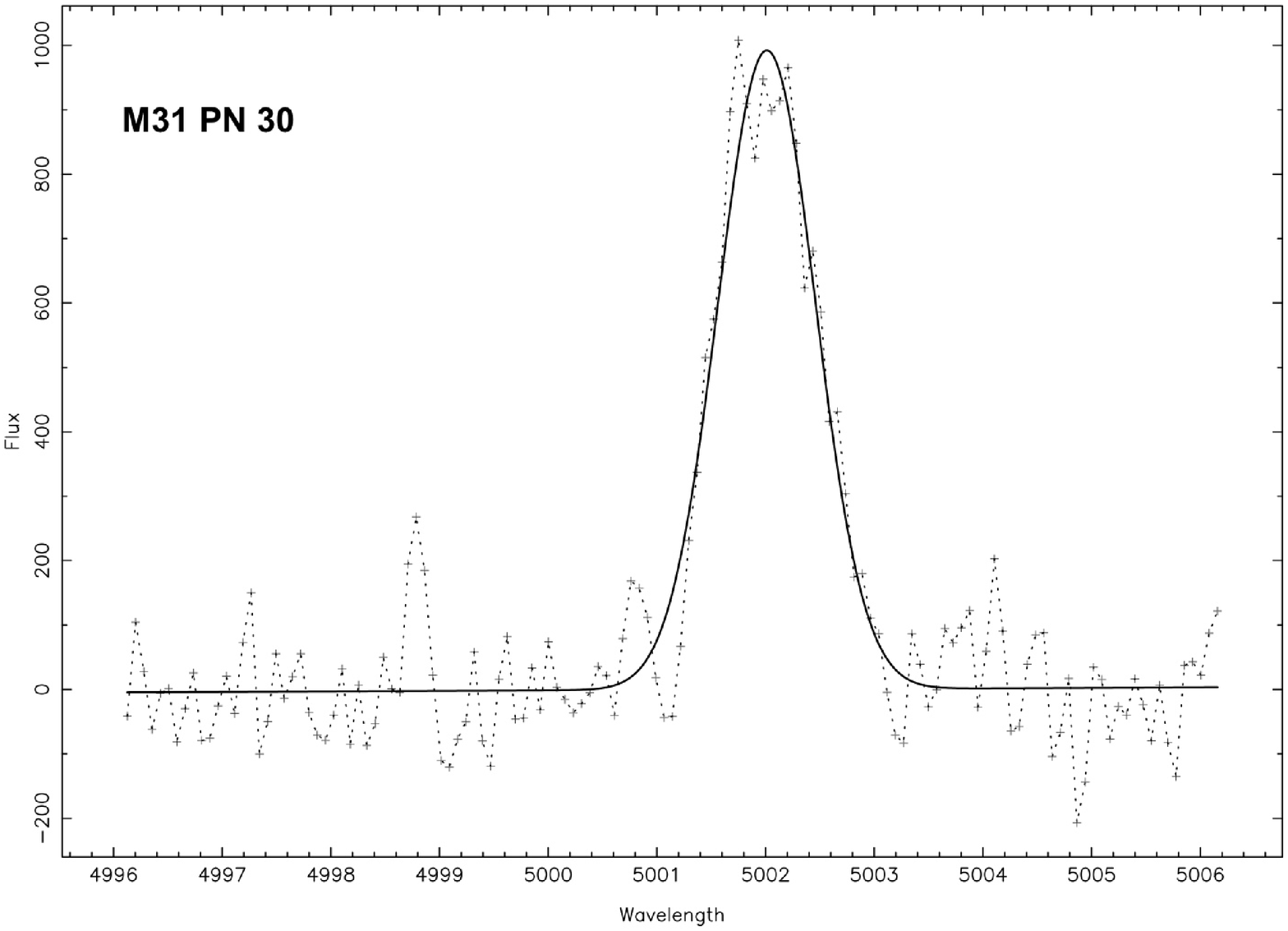} 
  \includegraphics[width=\columnwidth]{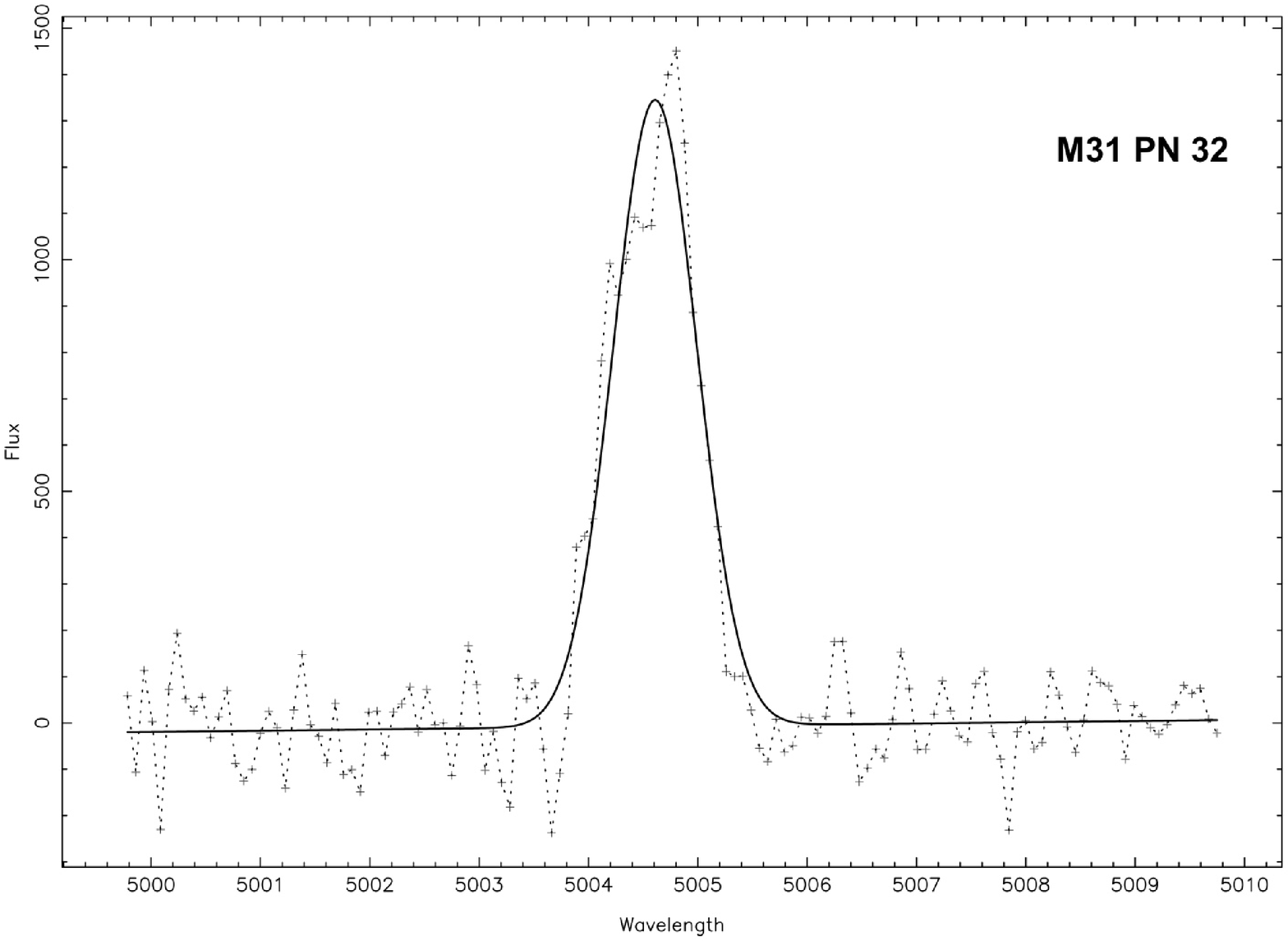} 
  \includegraphics[width=\columnwidth]{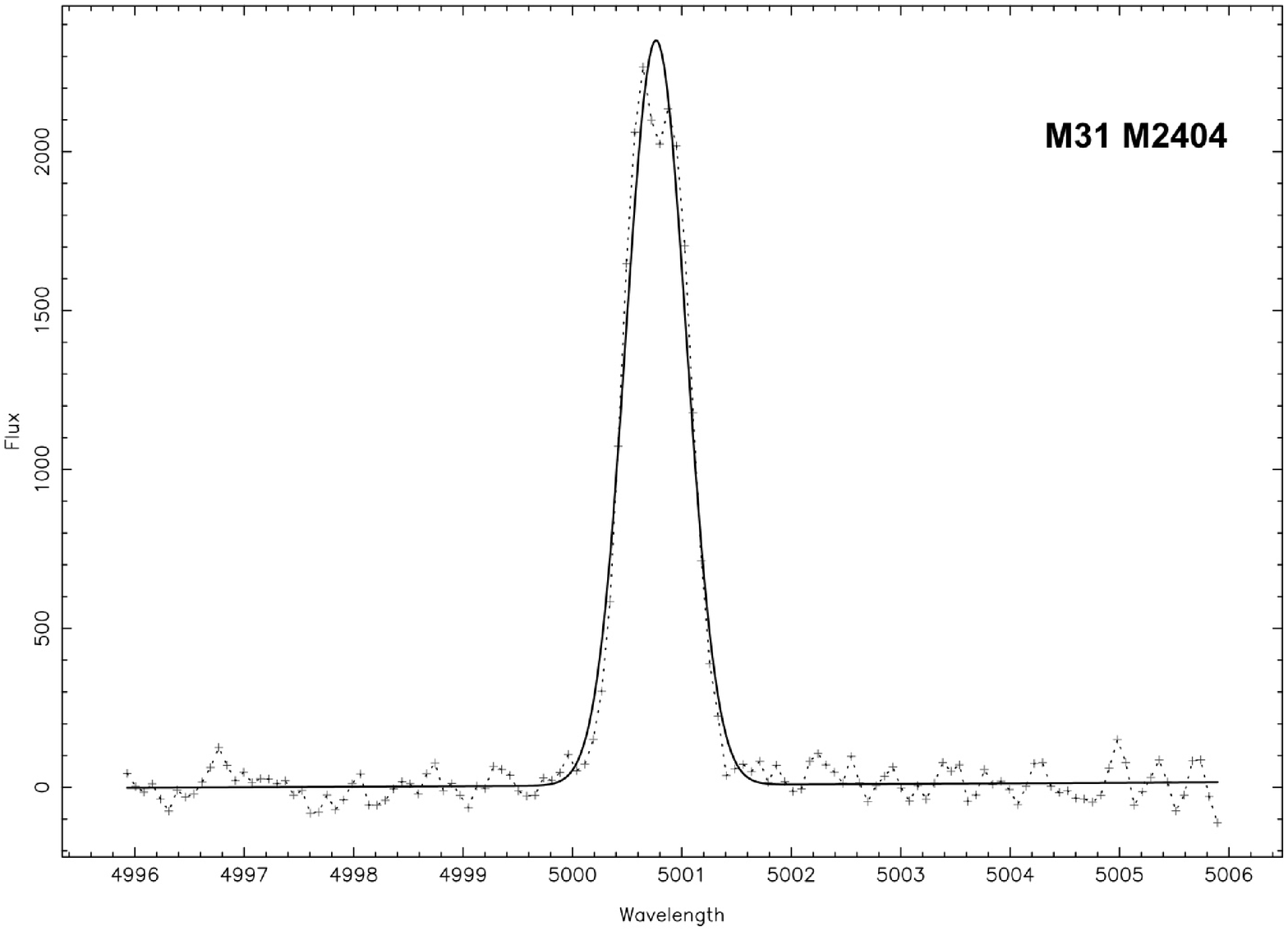} 
  \includegraphics[width=\columnwidth]{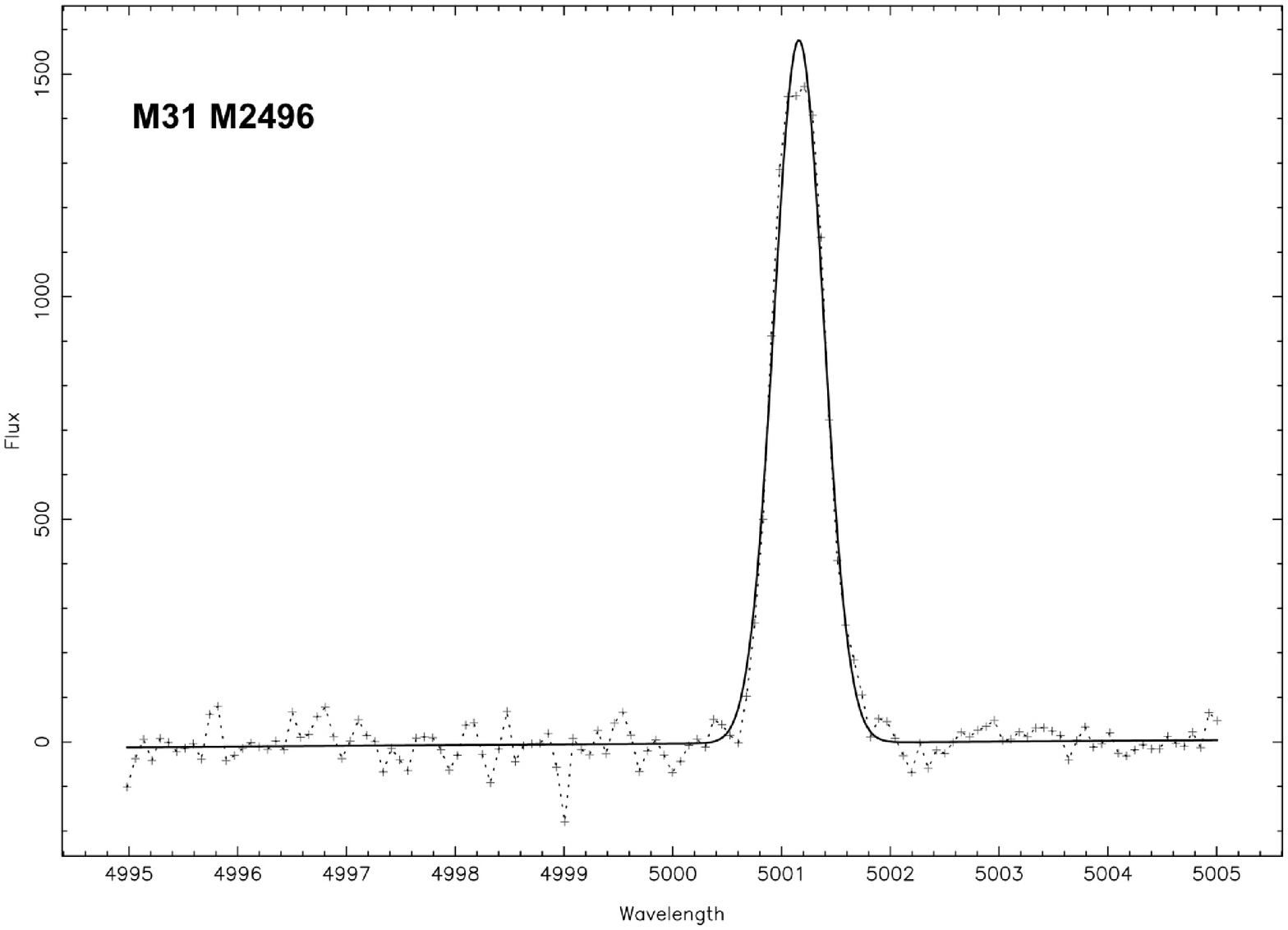} 
\end{center}
  \caption{The line profiles for four planetary nebulae in M31 are shown.  PN 30 and PN 32 belong to the bulge, M2404 to the disk, and M2496 to the halo.  Deviations of the data about the Gaussian fit are insignificant, even though these are among the profiles with the best S/N in M31 (PN 30 excepted).  See Fig. \ref{fig_line_prof_gallery1} for further details.}
\label{fig_line_prof_gallery3}
\end{figure*}

The line profiles of extragalactic planetary nebulae usually cannot be distinguished statistically from a Gaussian shape because of their limited S/N \citep[e.g.,][]{dopitaetal1985, dopitaetal1988, arnaboldietal2008}.  We analyzed the one-dimensional line profiles with a locally-implemented software package \citep[INTENS;][]{mccalletal1985} to determine the radial velocity, flux, and profile width (FWHM; full width at half maximum intensity) as well as the uncertainties ($1 \sigma$) in these parameters.  INTENS fits the emission line profile with a sampled Gaussian function and models the continuum as a straight line.  Thus, this analysis assumes that the lines have a Gaussian shape and that they are superposed on a flat continuum.  Figs. \ref{fig_line_prof_gallery1}-\ref{fig_line_prof_gallery3} present examples of the line profiles.  Excepting the planetary nebulae in the Fornax and Sagittarius dwarf spheroidals, the line profiles in Figs. \ref{fig_line_prof_gallery1}-\ref{fig_line_prof_gallery3} do not deviate significantly from a Gaussian shape, even though they are among our profiles with the highest S/N.  For the [\ion{O}{3}]$\lambda$5007 observations, we typically fit a 10\AA\ spectral interval more or less centered on the emission line, though as little as 6\AA\ was used in some cases to avoid cosmic rays.  For the H$\alpha$ observations, a 15\AA\ interval was used.  For the observations of StWr 2-21 and Wray 16-423 in the light of H$\alpha$, the \ion{He}{2}\,$\lambda$6560 line is also present, so we fit both lines simultaneously, but assuming  that the widths of both lines are identical.  

For each object we observed, Tables \ref{tab_o3}-\ref{tab_cHII} present the flux, the observed line width, the heliocentric radial velocity (from IRAF's rvcorrect), \emph{half} of the intrinsic line width (see below, Eq. \ref{eq_sigma}), and the corresponding uncertainties in all of these quantities.  The uncertainties are formal, one-sigma uncertainties and pertain to the final (sometimes summed) spectra.

As already noted (\S \ref{observations}), we did not attempt a flux calibration.  The vast majority of the objects in our catalogue (Table \ref{observations} contains the few exceptions) already have published photometry that is undoubtedly more precise that our spectroscopic fluxes would have been.  In practice, our fluxes provided a guide as to the need for additional observations.  We found that the accumulation of 4000--5000 counts was sufficient to provide adequate line profiles for analysis, provided that the line profiles were not too wide.  These fluxes allow the reader to judge the quality of the line profiles independently of the uncertainties attached to the line widths and central wavelengths.   

The radial velocities and the observed line widths in Tables \ref{tab_o3}-\ref{tab_cHII} are obtained as a result of fitting a Gaussian function to the line profile.  Their \emph{formal} uncertainties will depend upon the S/N and shape of the line profile.  Given line shapes that are nearly Gaussian and reasonable S/N, we commonly obtain formal uncertainties in the line widths and radial velocities that are substantially smaller than the instrument's absolute internal precision (\S \ref{observations}).  While these formal uncertainties are relevant for the line widths (observed and intrinsic), it is no surprise that the uncertainties in the radial velocities are dominated by systematic calibration issues (\S \ref{radial_velocities}).  

To compute the intrinsic line widths, the observed line widths are corrected for instrumental, Doppler/thermal, and fine structure broadening, supposing that all add in quadrature, i.e.,

\begin{equation}
\sigma^2_{obs} = \sigma^2_{true} + \sigma^2_{inst} + \sigma^2_{th} + \sigma^2_{fs}\ .\label{eq_sigma}
\end{equation}

\noindent where $\sigma_{true}$ is the intrinsic line width resulting from the kinematics of the planetary nebula while $\sigma_{inst}$, $\sigma_{th}$, and $\sigma_{fs}$ represent the instrumental, thermal, and fine structure broadening, respectively.  The instrumental profile is very nearly Gaussian with the pixel binning used and has a measured FWHM of 2.5-2.7 pixels, so we adopted a FWHM of 2.6 pixels for all objects ($\sim 11$\,km/s FWHM).  We compute the thermal broadening from the usual formula \citep[][eq. 2-243]{lang1980}, adopting rest wavelengths of 6562.83\AA\ and 5006.85\AA\ for H$\alpha$ and [\ion{O}{3}]$\lambda 5007$, respectively, assuming an electron temperature of $10^4$\,K and zero turbulent velocity.  The only exceptions were the planetary nebulae in the Fornax and Sagittarius dwarf spheroidal galaxies, for which the electron temperatures were adopted from the literature \citep{dudziaketal2000, zijlstraetal2006, kniazevetal2007}.  The resulting thermal broadening (FWHM) at $10^4$\,K amounts to 0.47\AA\ (21.4\,km/s) and 0.089\AA\ (5.3\,km/s) for H$\alpha$ and [\ion{O}{3}]$\lambda 5007$, respectively.  The fine structure broadening \citep{meaburn1970}, $\sigma_{fs}$, was taken to be 3.199\,km/s (FWHM 7.53\,km/s) for H$\alpha$ and zero for [\ion{O}{3}]$\lambda 5007$ \citep{garciadiazetal2008}.  

For a homogeneous, spherical, thin shell that fits entirely within the spectrometer slit, the resulting corrected line width,

\begin{equation}
\Delta V = 2.3556\sigma_{true}, 
\end{equation}

\noindent is twice the expansion velocity.   Only StWr 2-21 in the Sagittarius dwarf spheroidal is larger than our  slit \citep[][]{zijlstraetal2006}.  Departures from these assumptions in real objects affect the interpretation of $\Delta V$ \citep{schonberneretal2005}.  The ionized shells of planetary nebulae are neither infinitely thin, uniformly expanding, nor perfectly spherically symmetric.  Furthermore, velocity gradients within the ionized shell are relatively common \citep[see][for early examples]{wilson1950}.  In practice, it is clear that $\Delta V$ will be approximately twice the typical luminosity-weighted projected velocity at which the mass of the $\mathrm O^{2+}$ zone is flowing away from the central star.  The pattern velocity of the shock or ionization front that corresponds to the expansion velocity is substantially larger \citep{martenetal1993, steffenetal1997, schonberneretal2005}.  Given that, we adopt 

\begin{equation}
\Delta V_{0.5} = 0.5\Delta V = 1.1778\sigma_{true}, 
\end{equation}

\noindent as our measure of the kinematics of the matter in the $\mathrm O^{2+}$ zone.  It is this line width that is tabulated in Tables \ref{tab_o3}-\ref{tab_cHII}.  Since $\Delta V_{0.5}$ is obtained by fitting a Gaussian function, the line width at any other intensity may be derived from it.  \citet{richeretal2009} further discuss the limitations of this analysis.

\section{Results}\label{results}

\subsection{Line Profiles}

As expected, Figs. \ref{fig_line_prof_gallery1}-\ref{fig_line_prof_gallery3} demonstrate that the spatially-integrated line profiles for bright extragalactic planetary nebulae are Gaussian in shape, or nearly so, in agreement with previous results \citep{dopitaetal1985, dopitaetal1988, zijlstraetal2006, arnaboldietal2008}.  Our deepest profiles, however, show some additional structure.  Based upon observations of bright, compact planetary nebulae in the Milky Way bulge, \citet{richeretal2009} found that the Gaussian component represented more than 75\% of the [\ion{O}{3}]$\lambda 5007$ emission in 89\% of all cases.  Therefore, it is clear that the Gaussian component is an adequate representation of most of the line emission.  \citet{richeretal2009} also found that the line width of the Gaussian component was recovered reliably at the flux levels typical of bright extragalactic planetary nebulae.  

\subsection{Line Widths: H$\alpha$ Versus [\ion{O}{3}]$\lambda 5007$}

\begin{figure}[!t]
\begin{center}
  \includegraphics[width=1.0\columnwidth]{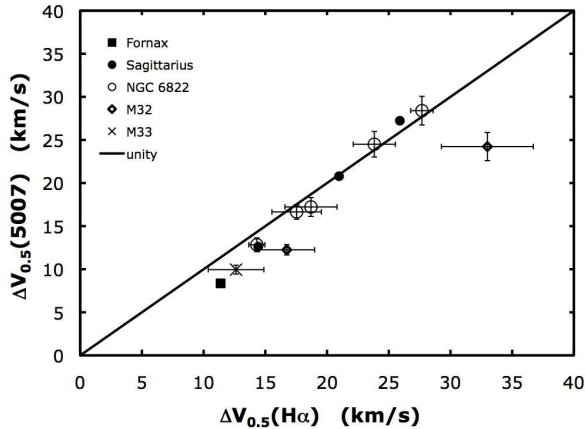}
\end{center}
  \caption{We compare the line widths measured in the lines of H$\alpha$ and [\ion{O}{3}]$\lambda 5007$.  The solid line indicates the locus of identical line widths.  The error bars show the formal uncertainty in line widths from the INTENS fit, which is smaller than the symbol for the planetary nebulae in Fornax and Sagittarius.  The intrinsic widths of the two lines are similar, as is also found for planetary nebulae in the Milky Way bulge \citep{richeretal2009}.}
\label{fig_dv_o3_ha}
\end{figure}

Fig. \ref{fig_dv_o3_ha} compares the H$\alpha$ and [\ion{O}{3}]$\lambda 5007$ line widths for those few objects for which both lines were observed.  As already noted by \citet{richeretal2009}, the line widths for the two lines are similar in bright planetary nebulae in the Milky Way bulge that were selected to mimic samples of bright extragalactic planetary nebulae in environments without star formation.  Taken with the result previously presented, the line width measured from the [\ion{O}{3}]$\lambda 5007$ line should be an adequate and reliable description of the kinematics of the entire ionized shell in bright extragalactic planetary nebulae.  

\subsection{Radial Velocities}\label{radial_velocities}

\begin{figure}[!t]
\begin{center}
  \includegraphics[width=1.0\columnwidth]{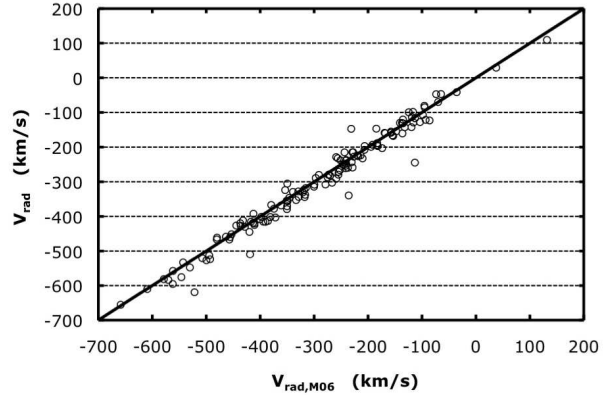}
\end{center}
  \caption{We compare our radial velocities with those measured by \citet[][M06]{merrettetal2006} for planetary nebulae in M31, M32, and NGC 205.  The solid line indicates the locus of identical radial velocities.  Our velocities are offset slightly, by -3.3\,km/s.}
\label{fig_compare_merrett}
\end{figure}

Although we made no effort to obtain precise absolute radial velocities, i.e., we observed no radial velocity standards, our heliocentric radial velocities turn out to be reasonably precise.  In computing the radial velocities, we adopt the same rest wavelengths as for the analysis of thermal broadening (\S \ref{analysis}).  The largest sample with which we can compare our radial velocities is that reported by \citet[][M06]{merrettetal2006} with which we have 136 objects in common in M31, M32, and NGC 205.  (Note that \citet{merrettetal2006} adopt their absolute velocity scale from \citet{hallidayetal2006}.) Fig. \ref{fig_compare_merrett} presents the correlation between the heliocentric radial velocities for the two data sets.  Clearly, the correlation is very good, with the exception of five objects (PN1, PN12, PN23, PN69, and PN190 in M31).  Excluding these objects, a linear least squares fit finds

\begin{equation}
V_{rad} = (1.005\pm 0.010)V_{M06} + (-3.3\pm 3.2)\,\mathrm{km/s}\,.
\end{equation}

\noindent  Thus, our heliocentric radial velocities differ only marginally from those of \citet{merrettetal2006}, with a possible offset of only -3.3\,km/s.  The dispersion about this relation is 16.7\,km/s.  Considering that \citet{merrettetal2006} claim an intrinsic dispersion of approximately 14\,km/s for their measurements, our radial velocities should then have an intrinsic dispersion of about 9\,km/s.  

\begin{figure}[!t]
\begin{center}
  \includegraphics[width=1.0\columnwidth]{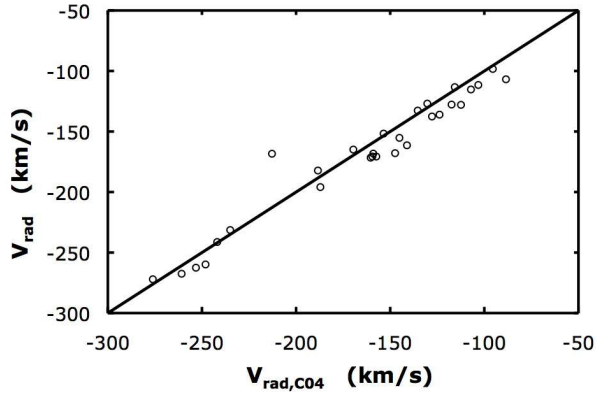}
\end{center}
  \caption{We compare our radial velocities with those measured by \citet[][C04]{ciardulloetal2004} for planetary nebulae in M33.  The solid line indicates the locus of identical radial velocities.  Our velocities are offset, by -13.8\,km/s.}
\label{fig_compare_ciardullo}
\end{figure}

We can also compare our heliocentric radial velocities for planetary nebulae in M33 with those of \citet[][C04]{ciardulloetal2004}.  Fig. \ref{fig_compare_ciardullo} presents the correlation for the 29 objects common to the two data sets, which is again very good, with M074 being the only deviant data point.  A linear least squares fit finds 

\begin{equation}
V_{rad} = (0.956\pm 0.027)V_{C04} + (-13.8\pm 4.7)\,\mathrm{km/s}\,.
\end{equation}

\noindent  Again, there is only a marginal difference, apart from an offset of -13.8\,km/s.  The dispersion about the above relation is 8.0\,km/s.  From Table 5 in \citet{ciardulloetal2004}, their typical uncertainty should be about 5.5\,km/s, implying that our radial velocities should have an intrinsic dispersion of about 5.8\,km/s.  

It is no coincidence that the intrinsic dispersion in our radial velocities is slightly less than the FWHM (11 km/s) we measure for the arc lamp.  Objects are centered in the slit via the spectrometer's imaging mode using short double exposures (sky and then slit) before the spectroscopic exposure.   While this guarantees that the object is centered in the slit during the acquisition exposure, small differential guiding errors between the acquisition and spectrosocpic exposures can result in it being displaced from the centre of the slit towards one side during the spectroscopic exposure.  Since the spectrometer resolves the slit used for these observations (\S \ref{observations}), the resulting radial velocities will be dispersed over a range somewhat less than the FWHM of the arc lines.  

We find slightly different offsets with respect to \citet{merrettetal2006} and \citet{ciardulloetal2004}.  This either reflects differences between their velocity scales or slight systematic changes in our radial velocity zero points from one observing run to the next.  Given that neither we nor \citet{ciardulloetal2004}, \citet{merrettetal2006}, or \citet{hallidayetal2006} observed any radial velocity standards, the small offsets noted are not surprising.  We therefore conclude that our radial velocities have an absolute precision on the order of 10\,km/s.  

\subsection{Line Width Versus Stellar Population}

A perusal of Tables \ref{tab_o3}-\ref{tab_Ha} reveals that the typical line widths, $\Delta V_{0.5}$, span a limited range in all galaxies.  Typically, $\Delta V_{0.5}$ is between 13 and 24\,km/s, though the total range extends from 6 to 30\,km/s.  Considering other similar data in the literature, the foregoing changes relatively little, though the example of SMP83 in the LMC demonstrates that occasionally larger line widths may be found among planetary nebulae within 2.5\,mag of the peak of the PNLF \citep{dopitaetal1985, dopitaetal1988, zijlstraetal2006, arnaboldietal2008}.

\begin{figure}[!t]
\begin{center}
  \includegraphics[width=1.0\columnwidth]{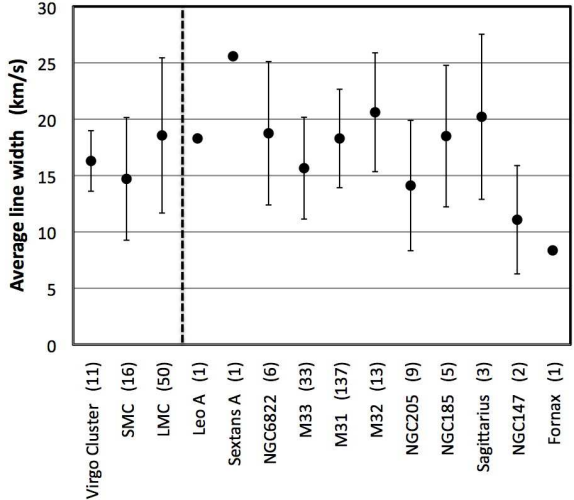}
\end{center}
  \caption{We compare the average [\ion{O}{3}]$\lambda$5007 line width from Table \ref{tab_o3}, $\Delta V_{0.5}$, found for the planetary nebulae in different galaxies.  The \lq\lq error bars" indicate the standard deviations of the line width distributions.  The number in parentheses indicates the number of planetary nebulae observed in each galaxy.  Clearly, the average line width does not vary much among different galaxies.  The data for the Magellanic Clouds are taken from \citet{dopitaetal1985, dopitaetal1988} for planetary nebulae within 2.5\,mag of the PNLF peak.  The data for the Virgo Cluster are taken from \citet{arnaboldietal2008}.}
\label{fig_line_width}
\end{figure}

Fig. \ref{fig_line_width} presents the average line width, $\Delta V_{0.5}$, observed for the planetary nebula populations in each of the galaxies in our sample.  The \lq\lq error bars" indicate the standard deviation of the $\Delta V_{0.5}$ distribution.  In this figure, the star-forming systems are on the left while those without active, on-going star formation are to the right.  M31 is a mixed system in this diagram, with on-going star formation in its disc, but with the bulge having ceased this activity long ago.  The data points for the Magellanic Clouds are taken from the data published by \citet{dopitaetal1985, dopitaetal1988} for planetary nebulae within 2.5\,mag of the PNLF peak, but using the same line width definition as for our data.  The data point for the Virgo Cluster is based upon the data published by \citet{arnaboldietal2008} for its \emph{intracluster} planetary nebulae.  Our only modification of their results was to correct for thermal broadening as we did for our data.  Obviously, the progenitor stellar population for these intracluster planetary nebulae is unknown.  

\begin{figure}[!t]
\begin{center}
  \includegraphics[width=1.0\columnwidth]{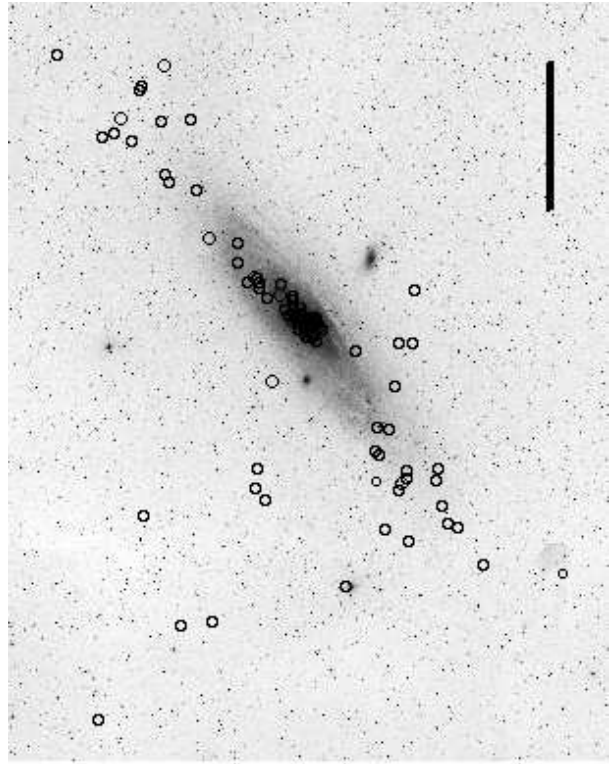}
\end{center}
  \caption{The planetary nebulae observed in M31 are shown overplotted on a DSS image of M31, M32, and NGC 205.  The bar is one degree in length.  North is up and east is to the left.}
\label{fig_M31_pne}
\end{figure}

\begin{figure}[!t]
\begin{center}
  \includegraphics[width=1.0\columnwidth]{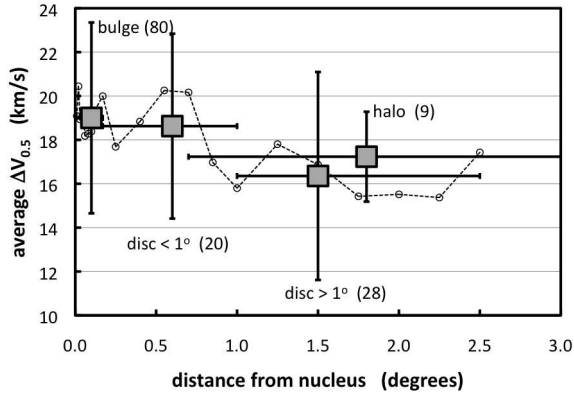}
\end{center}
  \caption{For the planetary nebulae in M31, we plot the average of $\Delta V_{0.5}$ for different samples as a function of distance from the nucleus.  The large symbols denote averages for the bulge, disc, and halo planetary nebulae (as defined in the text).  The numbers next to the labels indicate the number of objects in each sample.  The vertical error bars indicate the standard deviation within each group while the horizontal error bars denote the ratial extent.  The small symbols show the local average $\Delta V_{0.5}$ for bulge and disc planetary nebulae.  There is a clear tendency towards lower line widths at greater radial distances.}
\label{fig_M31_radial}
\end{figure}

The average line width ($\Delta V_{0.5}$) falls in a narrow range in Fig. \ref{fig_line_width}.  For all galaxies with at least three planetary nebulae observed, the average $\Delta V_{0.5}$ falls in the range of 14-21\,km/s.  There is no obvious difference between galaxies with and without on-going star formation.  This velocity range also includes the average line width (16.3\,km/s) for intracluster planetary nebulae in the Virgo cluster core \citep{arnaboldietal2008}.  

Likewise, the range of line widths ($\Delta V_{0.5}$) is also similar among galaxies.  There is a tendency to find a slightly larger range of line widths in the dwarf galaxies compared to M31 and M33.  If this tendency is statistically significant, it occurs for both dwarf irregulars and dwarf spheroidals, so it would appear to be due to the metallicity rather than the age of the progenitor stars.  The line widths for the intracluster planetary nebulae in the Virgo cluster have a very narrow distribution, perhaps a result of the effects studied by \citet{villaveretal2006}.

\subsection{M31}

We observed planetary nebulae in several stellar subsystems in M31 (bulge, disk, halo) as may be appreciated from Fig. \ref{fig_M31_pne}.  For the purposes of the analysis here, we assign planetary nebulae to a given subsystem based upon their position within the galaxy alone.  This will clearly not be perfect, but adopting a more complex scheme, such as including kinematic information, would not necessarily provide a perfect solution either.  We designate as halo objects the nine planetary nebulae lying more than $0.5^{\circ}$ from the major axis, since this implies an inclination-corrected disc radius $>2.2^{\circ}$, or $>30.3$\,kpc for our adopted distance of 789\,kpc \citep{leeetal2003}.  Note that these nine planetary nebulae all have inclination-corrected, projected disc radii in excess of $3.2^{\circ}$/44\,kpc \citep[adopting the scheme from][]{huchraetal1991}, so it is very unlikely that they are disc objects.  We designate as bulge planetary nebulae the 80 objects within the effective radius of bulge \citep[$0.17^{\circ}$:][]{walterboskennicutt1988}.  The remaining 48 planetary nebulae are assigned to the disk.  Note that M32-PN29 has a radial velocity indicating that it belongs to M31 (the disk in our scheme).  Our bulge-disc division will produce a bulge sample that is truly dominated by bulge planetary nebulae, but the disc sample will be contaminated by bulge objects near the cut-off radius.  

While Fig. \ref{fig_line_width} demonstrates that, to first order, the typical outflow velocity is not especially sensitive to either the metallicity or age of the progenitor stellar population, differences are noticeable within M31.  In Fig. \ref{fig_M31_radial}, we plot the average outflow velocity for the bulge, inner disc ($<1^{\circ}$), outer disc ($>1^{\circ}$), and halo samples as well as a local average for bulge and disc planetary nebulae.  Clearly, the line width, $\Delta V_{0.5}$, is lower for planetary at greater distances from the nucleus, though the inner disc and bulge have similar values.  On the other hand, the width of the velocity distribution does not vary with nuclear distance.  

The robustness of this trend can be tested via statistical tests.  The simplest comparison is to consider the distribution of line widths for planetary nebulae in the bulge and disk.  Applying the U-test to these subsamples \citep[][]{walljenkins2003}, there is only a 2.4\% probability of obtaining the two line width distributions by chance from a single progenitor population.  If the disk subsample is restricted to those objects at greater than $1^{\circ}$ from the nucleus, thereby minimizing the contamination of the disc sample by bulge objects, the probability of obtaining the two distributions by chance from a single parent population drops to 0.5\%.  Therefore, the trend of lower average line width ($\Delta V_{0.5}$) at greater radial distances is statistically significant.  

\begin{figure*}[!t]
\begin{center}
  \includegraphics[width=1.7\columnwidth]{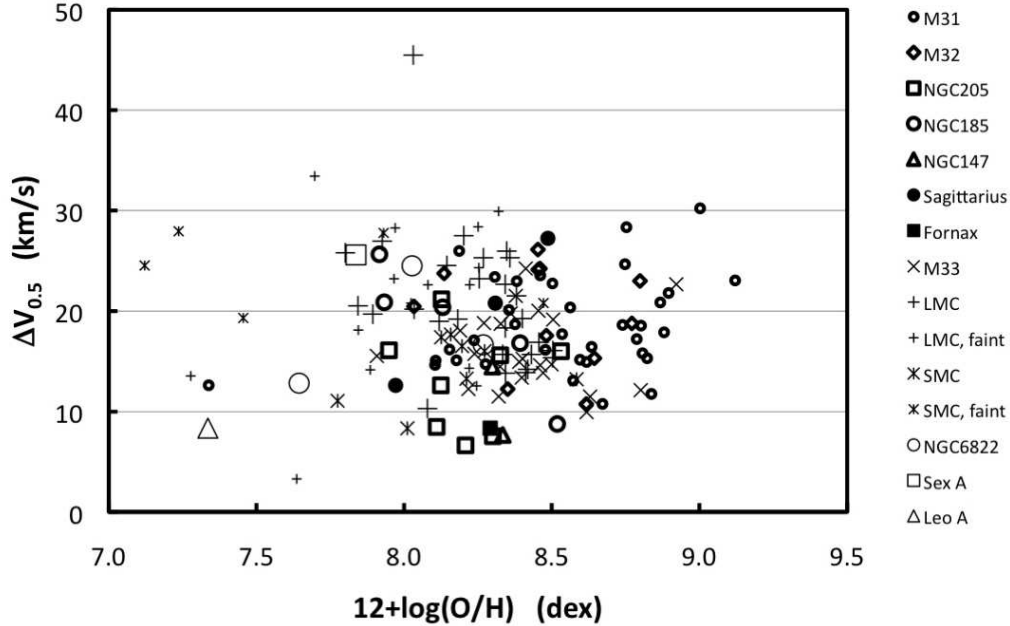}
\end{center}
  \caption{We plot the [\ion{O}{3}]$\lambda 5007$ line width as a function of the oxygen abundance for the planetary nebulae with low resolution spectra.  The symbols for galaxies without star formation are in bold or filled while the symbols for galaxies with ongoing star formation are light and unfilled.  Bright and faint planetary nebulae in the Magellanic Clouds are distinguished by large and small symbols, respectively.  For M31, the planetary nebulae come overwhelmingly from the bulge.  Galactic systems with and without star formation cover the abundance range from $7.9<12+\log(\mathrm O/\mathrm H)<8.5$, but planetary nebulae from the bulge of M31 dominate higher abundances.  There is no strong variation in line width with abundance or between galactic systems with and without star formation.  The oxygen abundances are taken from \citet{richermccall2008} and \citet{magrinietal2009}, the former adopting data from \citet{jacobyford1986, alleretal1987, barlow1987, monketal1987, woodetal1987, dopitaetal1988, meatheringhametal1988, henryetal1989, meatheringhamdopita1991a, meatheringhamdopita1991b, dopitameatheringham1991, vassiliadisetal1992, jacobykaler1993, leisydennefeld1996, walshetal1997, jacobyciardullo1999, richeretal1999, rothetal2004, kniazevetal2005, magrinietal2005, vanzeeetal2006, zijlstraetal2006, goncalvesetal2007, kniazevetal2007} and \citet{richermccall2007, richermccall2008}.}
\label{fig_dv_o3abun}
\end{figure*}

\begin{figure*}[!t]
\begin{center}
  \includegraphics[width=1.7\columnwidth]{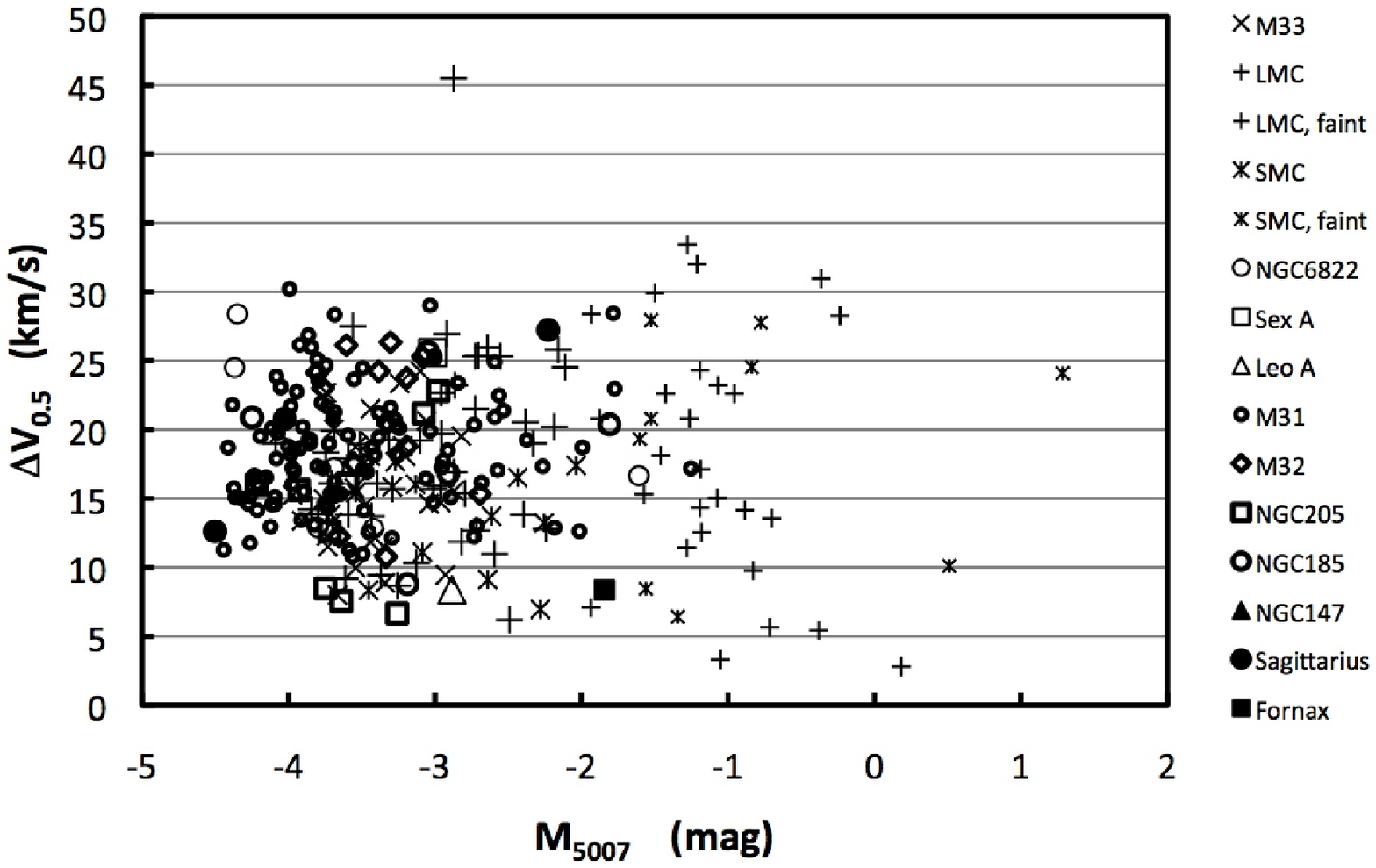}
\end{center}
  \caption{We plot the [\ion{O}{3}]$\lambda 5007$ line width as a function of the absolute [\ion{O}{3}]$\lambda 5007$ magnitude, $M_{5007}$, for those objects with appropriate data from the literature (for $M_{5007}$ references, see \S \ref{observations} for photometry and Fig. \ref{fig_dv_o3abun} for spectroscopy).  The distances are adopted from \citet{richermccall1995} and \citet{leeetal2003}, except for M33 and NGC 147, for which we adopt distance moduli from \citet{mcconnachieetal2005} since their distances (tip of red giant branch) for M31, NGC 185, and NGC 205 are consistent with those we adopt.  The symbols follow the same scheme as in Fig. \ref{fig_dv_o3abun}.  Note, however, that this figure includes many planetary nebulae from M31's disk.  As in Fig. \ref{fig_dv_o3abun}, no strong trend with line width is seen.  The only systematic effect is an increased dispersion for $M_{5007}>-2$\,mag.}
\label{fig_dv_M5007}
\end{figure*}

\begin{figure*}[!t]
\begin{center}
  \includegraphics[width=1.7\columnwidth]{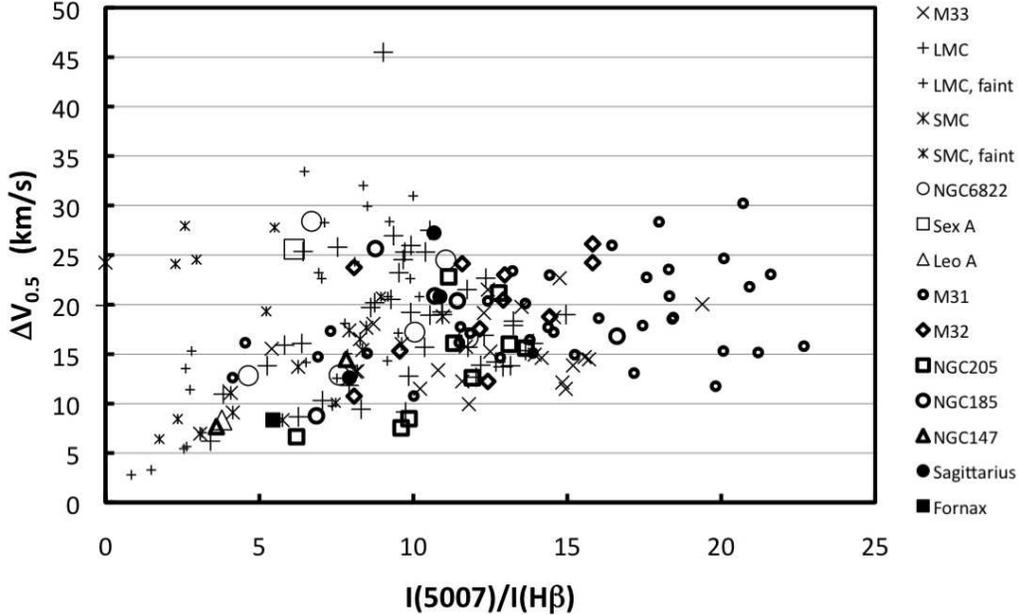}
\end{center}
  \caption{We plot the [\ion{O}{3}]$\lambda 5007$ line width as a function of the $I(5007)/I(\mathrm H\beta)$ ratio for the planetary nebulae with low resolution spectra (see the literature cited in Fig. \ref{fig_dv_o3abun}).  The symbols follow the same scheme as in Fig. \ref{fig_dv_o3abun}.  Here, most of the M31 planetary nebulae are from its bulge.  In part, the higher oxygen abundances for the planetary nebulae in M31's bulge allow them to achieve higher $I(5007)/I(\mathrm H\beta)$ ratios than for planetary nebulae elsewhere.  However, the planetary nebulae in M32 and the dwarf spheroidals tend to have systematically larger ratios than their counterparts in the dwarf irregulars, even though they have similar oxygen abundances, so nebular structure or the distribution of central star temperatures also likely play a role in elevating the $I(5007)/I(\mathrm H\beta)$ ratios for planetary nebulae in systems without star formation.}
\label{fig_dv_o3hbrat}
\end{figure*}

\begin{figure*}[!t]
\begin{center}
  \includegraphics[width=1.7\columnwidth]{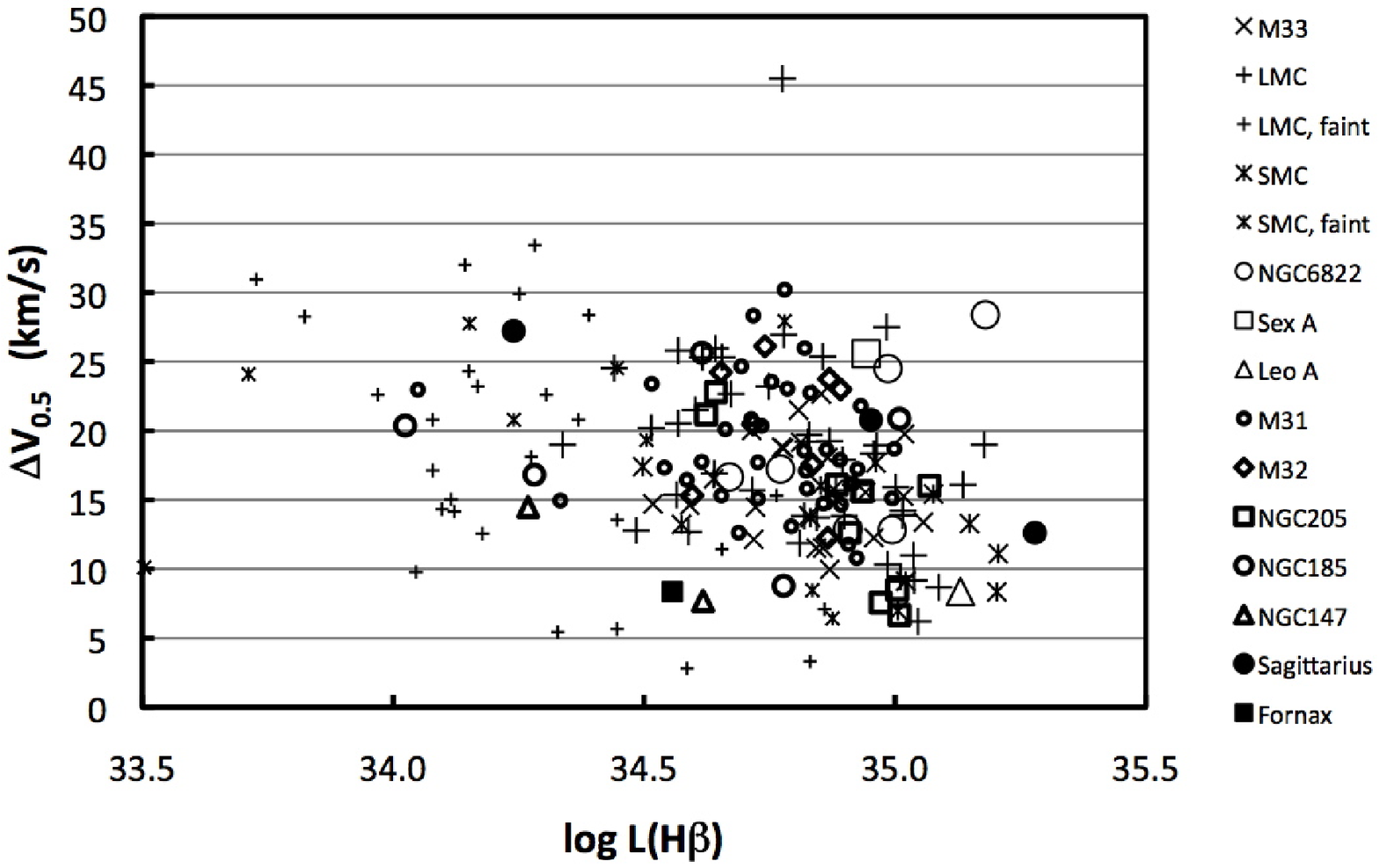}
\end{center}
  \caption{We plot the [\ion{O}{3}]$\lambda 5007$ line width as a function of the H$\beta$ luminosity of the planetary nebula for those objects with low resolution spectra from the literature.  The luminosities are based upon the distances and photometry from the references given in Fig. \ref{fig_dv_M5007} and \S2, respectively, and the [\ion{O}{3}]$\lambda 5007/\mathrm H\beta$ ratios from the literature cited in Fig. \ref{fig_dv_o3abun}.  The symbols follow the same scheme as in Fig. \ref{fig_dv_o3abun}.  Again, most of the M31 planetary nebulae are from its bulge.  At the brightest H$\beta$ luminosities there is a lack of the largest line widths.  Conversely, at the faintest H$\beta$ luminosities, the opposite occurs and there is a lack of the smallest line widths.}
\label{fig_dv_hblum}
\end{figure*}

\section{Discussion}\label{discussion}

Fig. \ref{fig_dv_o3_ha} indicates that the line widths in the lines of H$\alpha$ and [\ion{O}{3}]$\lambda 5007$ are similar, a result found for planetary nebulae in the Milky Way bulge \citep{richeretal2009}.  Hence, the line width observed for the brighter [\ion{O}{3}]$\lambda 5007$ line is a reasonable approximation of the line width for the entire ionized nebular shell.  
In all likelihood, this result arises because the intrinsically brightest planetary nebulae in the [\ion{O}{3}]$\lambda 5007$ line, such as those observed here, probably have O$^{2+}$ zones that occupy a large fraction of the volume of the ionized shell.  

Fig. \ref{fig_dv_o3_ha} also implies that, for planetary nebulae that are intrinsically bright in [\ion{O}{3}]$\lambda 5007$, ionization stratification \citep[e.g.,][]{wilson1950} does \emph{not} strongly bias the width of the [\ion{O}{3}]$\lambda 5007$ line.  Ionization stratification might explain the trend of systematically lower [\ion{O}{3}]$\lambda 5007$ line widths at small H$\alpha$ line widths, but even this effect is small.  Therefore, in spite of ionization stratification, for it undoubtedly must occur, the [\ion{O}{3}]$\lambda 5007$ line width is a good estimator of the line width for the entire ionized shell, as measured by H$\alpha$.  The large volume of the O$^{2+}$ zone plays an important role in establishing this result and is a product of our sample selection (high [\ion{O}{3}]$\lambda 5007$ luminosity).  For objects in other evolutionary phases, our result may not hold, as is commonly found in highly evolved objects where ionization stratification is regularly seen \citep[e.g.,][]{meaburnetal2008}.  

The individual planetary nebulae in Fig. \ref{fig_dv_o3_ha} arise from galaxies both with and without ongoing star formation.  Although the sample is small, there is no obvious indication that the relation between H$\alpha$ and [\ion{O}{3}]$\lambda 5007$ line widths differs between galaxies with and without ongoing star formation.  Therefore, it appears that the [\ion{O}{3}]$\lambda 5007$ line width is an equally good estimate of the line width for the entire ionized shell in the brightest planetary nebulae in all galaxies.

Interpreting Fig. \ref{fig_line_width} is challenging given present knowledge.  The hydrodynamical models of \citet{perinottoetal2004} and \citet{schonberneretal2005, schonberneretal2005b} appear to be the most relevant, given their experiments with the AGB mass-loss rate, wind velocity, and metallicity. \emph{Only} \citet{schonberneretal2005b} consider the effect of metallicity on hydrodynamical models of planetary nebula evolution, and they find a greater acceleration of the nebular shell at lower metallicity because of the higher electron temperature achieved in the ionized shell.  Both observation and the theory of dust-driven winds concur in that lower metallicity produces lower AGB wind velocities, though not necessarily lower mass-loss rates \citep[e.g.,][]{woodetal1992, marshalletal2004, mattssonetal2008, wachteretal2008, groenewegenetal2009, lagadecetal2010}.  Also, the circumstellar shells of low-mass OH/IR stars have lower expansion velocities than those surrounding their higher mass counterparts \citep{lewisetal1990}.  Finally, though theoretical studies commonly find that the final white dwarf mass depends upon the metallicity of the progenitor star \citep[e.g.,][]{dominguezetal1999, weidemann2000}, to date, observation has been unable to substantiate this prediction \citep[e.g.,][]{catalanetal2008}.  %

If we consider separately the star-forming or non-star-forming galaxies from Fig. \ref{fig_line_width}, each group approximates a metallicity sequence, though possibly with progenitor stellar populations of different ages.  There is no obvious correlation of the line width with metallicity in either group.  Since lower metallicity produces progenitor AGB stars with lower wind velocities, which translate into lower expansion velocities in the resulting planetary nebula \citep{perinottoetal2004}, other parameters being equal, the effect of metallicity must be  compensated.  \citet{schonberneretal2005b} suggest that the higher electron temperatures at lower metallicities, with their consequently larger thermal pressures, provide greater acceleration of the nebular shell.  Another possibility is that the progenitors of bright planetary nebulae with lower metallicity are slightly more massive, with their higher AGB wind velocities compensating for their lower metallicities.  Greater acceleration is the simpler solution.  Qualitatively, higher progenitor masses at lower metallicity allow the spectral differences noted in the past \citep{stasinskaetal1998, richer2006}, but the variation cannot be too great, in order to comply with the small range needed to maintain the similar chemical abundance ratios and the variation of the PNLF peak luminosity with metallicity that are observed \citep{ciardulloetal2004, richermccall2008}.

There is no clear difference between the line widths of planetary nebulae in galaxies with and without star formation in Fig. \ref{fig_line_width}.  Given the foregoing, the implication is that the stellar progenitors of bright planetary nebulae in all galaxies have similar masses, assuming that some mechanism exists to compensate for the lower velocities of the winds of AGB stars at lower metallicities.  If so, at a given metallicity, bright planetary nebulae represent stars formed at similar epochs in galaxies with and without star formation.  This may only be understood if the progenitor stars of bright planetary nebulae in all galaxies are of relatively low mass, since significant star formation stopped long ago (several Gyr) in many of the galaxies without star formation considered here.   Another option is that the denser ISM in star forming galaxies provides a greater impediment to the nebular outflow than does its counterpart in galaxies without star formation \citep[e.g.,][]{oeygarciasegura2004}.  If so, progenitor stars of higher masses would be feasible in galaxies with ongoing star formation.  Clearly, the issue must still be investigated in detail.

In galaxies with ongoing star formation, bright planetary nebulae usually have chemical abundances similar to those in the interstellar medium (ISM) in their host galaxies \citep{richermccall2007, magrinigoncalves2009, bresolinetal2010}.  While this may be interpreted to suggest that the progenitors of their planetary nebulae are relatively young because the composition of the ISM has not evolved significantly since their formation, the ease with which the ISM composition can be changed decreases as the metallicity increases (and only the galaxies Leo A and Sextans A have very low metallicity).  Thus, even in star-forming galaxies, the bright planetary nebulae need not arise from very recent star formation.  Indeed, the chemical composition of bright planetary nebulae is very similar in galaxies with and without star formation \citep{richermccall2008}, again suggesting that their progenitor stars were of similar masses.  

The planetary nebulae that descend from massive progenitor stars need not be among the brightest planetary nebulae because of their rapid evolution \citep[e.g.,][]{villaveretal2002}.  On the other hand, they can produce the planetary nebulae with the largest outflow velocities for the ionized shell \citep[e.g.,][]{villaveretal2002, perinottoetal2004}.  SMP83 in the LMC, the high point in Figs. \ref{fig_dv_o3abun}-\ref{fig_dv_hblum} may be an example of a planetary nebula descended from a relatively massive progenitor that is also among the brightest.

Fig. \ref{fig_line_width} can therefore be understood if two conditions are met.  First, some mechanism must exist that compensates the slower winds of low metallicity AGB stars.  The greater thermal pressure resulting from ionization is the most natural explanation \citep{schonberneretal2005b}, though it is not the only possibility.  Second, the masses of the progenitor stars of the brightest planetary nebulae in all galaxies should not differ too much, though some variation is feasible and even necessary to explain the differences in their low resolution spectra \citep{stasinskaetal1998, richer2006}.

The trend of line width ($\Delta V_{0.5}$) within M31 agrees with these results (Fig. \ref{fig_M31_radial}).  The chemical abundances in young stars and \ion{H}{2} regions indicate that the disc has a shallow oxygen abundance gradient \citep[e.g.,][]{dennefeldkunth1981, blairetal1982, vennetal2000, trundleetal2002}, with the oxygen abundance falling by less than a factor of 3 between nuclear distances of $0.5^{\circ}$ and $2^{\circ}$.  Since Fig. \ref{fig_M31_radial} implies that the line width decreases only slightly as metallicity decreases, it is again necessary to invoke some mechanism to compensate for the drop in AGB wind velocity at lower metallicity.%

In Fig. \ref{fig_dv_o3abun}, we plot the [\ion{O}{3}]$\lambda 5007$ line width as a function of the oxygen abundance.  Clearly, the [\ion{O}{3}]$\lambda 5007$ line width does not vary strongly as a function of oxygen abundance.  There may be greater dispersion among line widths for abundances below $12+\log(\mathrm O/\mathrm H)\sim 7.8$\,dex.  Likewise, there may be a deficit of narrow line widths above $12+\log(\mathrm O/\mathrm H)\sim 8.7$\,dex.  However, the statistics are poor in both cases.

In Fig. \ref{fig_dv_M5007}, we plot the line width as a function of the absolute [\ion{O}{3}]$\lambda 5007$ magnitude, $M_{5007}$.  Again, there is no clear variation of the line width with $M_{5007}$, nor is there any obvious difference between systems with and without ongoing star formation.  It is unfortunate that there are not more data for fainter planetary nebulae ($M_{5007} > -2$\,mag), especially for systems without ongoing star formation, as it would be useful to determine whether the increased dispersion observed for fainter planetary nebulae in the Magellanic Clouds is a more general result (Fig. \ref{fig_dv_M5007}).

When the line width is plotted as a function of the $I(5007)/I(\mathrm H\beta)$ ratio, there is a systematic difference in the distribution of line widths for the planetary nebulae from \emph{dwarf} galaxies with and without star formation.  In Fig. \ref{fig_dv_o3hbrat}, there is an offset between the planetary nebulae in systems with and without ongoing star formation with those from galaxies without star formation having larger $I(5007)/I(\mathrm H\beta)$ ratios at a given line width.  Based upon the U-test \citep{walljenkins2003}, the difference between galaxies with and without star formation is significant at the 90\% level when only bright planetary nebulae in \emph{dwarf galaxies} are included.  Since this happens even though the dwarf galaxies with and without star formation have similar oxygen abundances (Fig. \ref{fig_dv_o3abun}), it is not an abundance effect, so it presumably arises as a result of differences in ionization structure or in the distribution of central star temperatures.  The slight difference in progenitor masses proposed to explain Fig. \ref{fig_line_width} could conceivably produce this difference.

In Fig. \ref{fig_dv_o3hbrat}, the planetary nebulae in M31 extend to much higher $I(5007)/I(\mathrm H\beta)$ ratios than their counterparts in other galaxies.  These are the most oxygen-rich planetary nebulae and simulations indicate that the $I(5007)/I(\mathrm H\beta)$ ratio varies as the square root of the oxygen abundance \citep[][]{dopitaetal1992,mendezetal2005}.  However, the difference between the planetary nebulae in M31 and their counterparts in other galaxies is greater than this.  The median value of the oxygen abundances for planetary nebulae in M31 differs from that for the planetary nebulae in the dwarf galaxies by less than a factor of two, but the difference in $I(5007)/I(\mathrm H\beta)$ ratios is greater than 50\%, which suggests that the change in ionization structure may accentuate at higher oxygen abundances.   

The most obvious trend that we find as a function of line width is with H$\beta$ luminosity (Fig. \ref{fig_dv_hblum}).  The line widths define a broad swath in this diagram, but there are deficits of both large line widths at the highest H$\beta$ luminosities and narrow line widths at the lowest H$\beta$ luminosities.  These deficits, especially that at high H$\beta$ luminosity, are not likely to be a selection effect, since young planetary nebulae rapidly achieve their maximum luminosity \citep{schonberneretal2007}, so they should be found on the right side of Fig. \ref{fig_dv_hblum}.  If there exist planetary nebulae that have very low H$\beta$ luminosities when young (with low [\ion{O}{3}]$\lambda 5007/\mathrm H\beta$), they could conceivably help fill in the deficit of low line widths at low H$\beta$ luminosity.

The existence of a trend of line width with H$\beta$ luminosity (Fig. \ref{fig_dv_hblum}), but not with $M_{5007}$ magnitude (Fig. \ref{fig_dv_M5007}), has important implications for the temporal evolution of bright planetary nebulae.  Objects may be intrinsically bright in [\ion{O}{3}]$\lambda 5007$ if the $I(5007)/I(\mathrm H\beta)$ ratio is low or modest and the H$\beta$ luminosity is high, or if the $I(5007)/I(\mathrm H\beta)$ ratio is high and the H$\beta$ luminosity is at least modest.  Thus, on average, [\ion{O}{3}]$\lambda 5007$-bright planetary nebulae fade monotonically in the light of H$\beta$.  On the other hand, their evolution in [\ion{O}{3}]$\lambda 5007$ luminosity may be double-valued.  In order to reproduce the tendency found between line width and H$\beta$ luminosity, the line width evolution should also be monotonic, on average, increasing as the H$\beta$ luminosity decreases.  Recent hydrodynamical models concur with these three evolutionary trends, at least for stellar progenitors of lower masses \citep{villaveretal2002, perinottoetal2004, schonberneretal2005, schonberneretal2007}.

Supposing that Fig. \ref{fig_dv_hblum} represents an evolutionary sequence, it offers insight into the positions of Fornax, NGC 147, and NGC 205 in Fig. \ref{fig_line_width}.  For the planetary nebula in Fornax as well as PN04 in NGC 147 and PN01, PN07, and PN09 in NGC 205, low-resolution spectra find very weak \ion{He}{2}~$\lambda$4686, indicating that their central stars have not yet become very hot  \citep{goncalvesetal2007, kniazevetal2007, richermccall2008}.  These two lines of evidence then imply that we observe these objects early in the planetary nebula phase and that, for Fornax and NGC 147, they bias their positions in Fig. \ref{fig_line_width} to unusually low values.  

Metallicity likely plays a role in the tendency to find a larger range of line widths ($\Delta V_{0.5}$) in the dwarf galaxies as compared to M31 and M33.  The dwarf galaxies have lower metallicities and the peak luminosity of the PNLF is lower at lower metallicity \citep{ciardulloetal2002}.  This might allow progenitor stars spanning a wider range of masses to contribute to the bright end of the PNLF at lower metallicity.  Alternatively, planetary nebulae in a wider range of evolutionary states might contribute to the peak of the PNLF (Fig. \ref{fig_dv_hblum}).  Either possibility would produce a wider range in line width at lower metallicity.

Hydrodynamical models have consistently predicted that the velocities of nebular shells should increase with time for planetary nebulae arising from progenitors of modest masses \citep[$M \lesssim 2.5\,M_{\odot}$; e.g.,][also see previous paragraph]{schmidtvoigtkoppen1987, kahnbreitschwerdt1990, mellema1994}.  \citet{dopitaetal1985, dopitaetal1988} first proposed such an evolutionary scheme for the kinematics of planetary nebulae in the Magellanic Clouds.  \citet{richeretal2008, richeretal2010} have found similar evolution for the bright planetary nebulae in the Milky Way bulge.  Fig. \ref{fig_dv_hblum} indicates that the acceleration of nebular shells is a general phenomenon, at least for the intrinsically brightest planetary nebulae (in the light of [\ion{O}{3}]$\lambda 5007$).  Thus, there is now excellent qualitative agreement between theory and observation regarding the ionized shell's kinematic evolution for intrinsically bright planetary nebulae: with time, the shells are accelerated.  The central star's important effect upon the nebular shell is clear.  A more quantitative comparison must await the arrival of spatially-integrated line profiles from hydrodynamical models that may be directly compared with observations.  

For the PNLF to be a useful distance indicator, it must not be sensitive to the underlying stellar populations, a result found early in its use  \citep{ciardulloetal1989b}. \citet{merrettetal2006} confirm this, finding that the PNLF does not vary significantly with position in M31.  Meanwhile, \citet{richermccall2008} find that the progenitors of bright planetary nebulae in all galaxies have undergone similar chemical enrichment processes.  These results argue that the progenitor stellar populations of intrinsically bright planetary nebulae are similar in all galaxies, though they are clearly not identical \citep[e.g.,][]{stasinskaetal1998, ciardulloetal2004}.  Likewise, the similar kinematics we find for the brightest planetary nebulae in all galaxies would appear to argue that there is not a large range in the masses of the stellar progenitors of these objects in different galaxies, whether star-forming or not.  Their internal kinematics, therefore, add a third argument favoring progenitor stellar populations for the brightest planetary nebulae that are similar in all galaxies.  

\section{Conclusions}\label{conclusions}

We present kinematic data for 211 bright extragalactic planetary nebulae in 11 Local Group galaxies.  We present line profiles in the line of [\ion{O}{3}]$\lambda 5007$ for all objects and in the H$\alpha$ line for a small minority.  At the signal-to-noise of our data, the line profiles are Gaussian, or nearly so, in all cases.  The intrinsic line widths in [\ion{O}{3}]$\lambda 5007$ and H$\alpha$ are also similar.  Thus, the line widths in [\ion{O}{3}]$\lambda 5007$ are an adequate description of the kinematics of most of the matter in the entire ionized shell in these objects.

We find that the average line width for the bright planetary nebulae in all galaxies are similar (where we observed at least three objects).  Given our current understanding, this result implies that the progenitors of bright planetary nebulae at lower metallicity are either slightly more massive or that the higher plasma temperatures at lower metallicities produce larger acceleration of the nebular shell.  The approximate constancy of the line width also implies that there cannot be a large difference in the masses of the progenitors of bright planetary nebulae among galaxies, whether star-forming or not, though small variations are possible.  Within M31, the line width decreases with increasing distance from the nucleus, in agreement with these deductions.  

As a general rule, the range of line widths is larger in dwarf galaxies than in M31 or M33.  Presumably, this arises because, at lower metallicity, either a larger range of progenitor masses contribute bright planetary nebulae or that bright planetary nebulae encompass a wider range of evolutionary states. On the other hand, there is no obvious correlation between the line width in [\ion{O}{3}]$\lambda 5007$ and either the oxygen abundance or the absolute [\ion{O}{3}]$\lambda 5007$ magnitude, $M_{5007}$.  

The most significant correlation is between line width and the H$\beta$ luminosity.  Since [\ion{O}{3}]$\lambda 5007$-bright planetary nebulae represent a monotonic fading sequence in the light of H$\beta$, this correlation implies that the ionized shells of bright planetary nebulae are accelerated during their early evolution, independently of the age or metallicity of the progenitor stellar population.  Theory has long predicted this result, so our data are consistent with its general validity.

More generally, our measurements of the kinematics of bright planetary nebulae are in good accord with the results derived from photometry and low resolution spectroscopy.  All three lines of reasoning require that intrinsically bright planetary nebulae descend from progenitor stellar populations spanning a relatively small range of masses.  

We thank the telescope operators during our runs, Gabriel Garc\'\i a, Gustavo Melgoza, Salvador Monrroy, and Felipe Montalvo, for their help.  We thank M. Arnaboldi for kindly providing the line widths for the planetary nebulae in the Virgo Cluster that are shown in Fig. \ref{fig_line_width}.  We thank the referee for helpful comments and suggestions.  We gratefully acknowledge financial support throughout this project from CONACyT grants 37214, 43121, 49447, and 82066 and from DGAPA-UNAM grants 108406-2, 108506-2, 112103, and 116908-3.  We acknowledge the use of NASA's \emph{SkyView} facility (http://skyview.gsfc.nasa.gov) located at NASA Goddard Space Flight Center in generating Fig. \ref{fig_M31_pne}.

\clearpage\onecolumn
\tablecols{7}
\tabcaption{[\ion{O}{3}]$\lambda$5007 data for extragalactic planetary nebulae}

\def\ColumnHeaders{
Galaxy & Object & Flux\ $^{\mathrm a}$ & FWHM & $V_{helio}\ ^{\mathrm b}$ & $\Delta V_{0.5}$ & Run \\      
       &        & $(10^3\,\mathrm{ADU})$ & (\AA) & (km/s) & (km/s) &
}

\begin{longtable}{llccccl}
  \toprule
  \ColumnHeaders\\ \midrule
  \endfirsthead
  
  \tabcaptioncontinued
  \toprule
  \ColumnHeaders\\ \midrule
  \endhead
  
  \bottomrule
  \endfoot

  \bottomrule
  \multicolumn{7}{l}{$^{\mathrm a}$\ The fluxes are not calibrated.  See \S \ref{observations} and \S \ref{analysis}.}\hfill\\
  \multicolumn{7}{l}{$^{\mathrm b}$\ The uncertainties listed are formal uncertainties.  See \S \ref{radial_velocities}.\hfill}
  \endlastfoot
Fornax      & PN1        & $  40.5\pm 0.4 $ & $ 0.353\pm 0.004 $ & $   51.2\pm 0.1 $ & $  8.4\pm 0.1 $ & 2004nov \\
Leo A       & PN1        & $  16.5\pm 0.4 $ & $ 0.65\pm 0.02   $ & $   32.4\pm 0.4 $ & $ 18.3\pm 0.5 $ & 2004jun, 2004nov \\
M31         & f08n04     & $   6.9\pm 0.2 $ & $ 0.53\pm 0.02   $ & $ -381.7\pm 0.5 $ & $ 14.6\pm 0.6 $ & 2005sep \\
M31         & f08n05     & $  10.2\pm 0.3 $ & $ 0.79\pm 0.03   $ & $ -148.0\pm 0.7 $ & $ 22.7\pm 0.8 $ & 2004nov \\
M31         & f08n08     & $   7.0\pm 0.2 $ & $ 0.67\pm 0.02   $ & $ -229.2\pm 0.5 $ & $ 18.9\pm 0.6 $ & 2005sep \\
M31         & f08n09     & $   8.1\pm 0.2 $ & $ 0.424\pm 0.009 $ & $  -54.5\pm 0.2 $ & $ 10.9\pm 0.3 $ & 2004nov \\
M31         & f08n10     & $   7.8\pm 0.3 $ & $ 0.85\pm 0.03   $ & $  -50.9\pm 0.8 $ & $ 24.6\pm 0.9 $ & 2005sep \\
M31         & f15n12002  & $   9.2\pm 0.2 $ & $ 0.50\pm 0.01   $ & $  -90.8\pm 0.3 $ & $ 13.6\pm 0.3 $ & 2005sep \\
M31         & f29n2065   & $   4.2\pm 0.2 $ & $ 0.76\pm 0.04   $ & $ -474\pm 1     $ & $ 22\pm 1     $ & 2007aug \\
M31         & f29n9178   & $   5.5\pm 0.1 $ & $ 0.52\pm 0.02   $ & $ -524.1\pm 0.4 $ & $ 14.1\pm 0.5 $ & 2007aug \\
M31         & f32n1      & $   5.9\pm 0.2 $ & $ 0.59\pm 0.03   $ & $ -371.8\pm 0.6 $ & $ 16.4\pm 0.7 $ & 2004nov \\
M31         & fjchp51    & $   3.7\pm 0.2 $ & $ 0.54\pm 0.03   $ & $ -503.0\pm 0.8 $ & $ 15\pm 1     $ & 2007jan \\
M31         & fjchp57    & $   5.2\pm 0.2 $ & $ 0.54\pm 0.03   $ & $ -141.9\pm 0.7 $ & $ 14.7\pm 0.8 $ & 2007aug \\
M31         & M0050      & $  13.7\pm 0.2 $ & $ 0.549\pm 0.008 $ & $  -35.7\pm 0.2 $ & $ 15.1\pm 0.2 $ & 2005sep \\
M31         & M0273      & $   8.9\pm 0.2 $ & $ 0.84\pm 0.02   $ & $ -132.0\pm 0.5 $ & $ 24.2\pm 0.7 $ & 2005sep \\
M31         & M0319      & $  14.5\pm 0.2 $ & $ 0.546\pm 0.009 $ & $ -137.2\pm 0.2 $ & $ 15.0\pm 0.3 $ & 2005sep \\
M31         & M0442      & $  12.8\pm 0.2 $ & $ 0.597\pm 0.009 $ & $  -86.0\pm 0.2 $ & $ 16.7\pm 0.3 $ & 2005sep \\
M31         & M1232      & $   2.9\pm 0.2 $ & $ 0.85\pm 0.06   $ & $ -174\pm 1     $ & $ 25\pm 2     $ & 2006sep \\
M31         & M1296      & $   1.6\pm 0.2 $ & $ 0.68\pm 0.08   $ & $  -90\pm 2     $ & $ 19\pm 2     $ & 2001sep \\
M31         & M1558      & $  12.2\pm 0.2 $ & $ 0.59\pm 0.01   $ & $ -349.7\pm 0.3 $ & $ 16.5\pm 0.4 $ & 2005sep \\
M31         & M1596      & $   8.0\pm 0.2 $ & $ 0.76\pm 0.02   $ & $ -374.5\pm 0.5 $ & $ 21.7\pm 0.7 $ & 2007aug \\
M31         & M1980      & $   7.5\pm 0.2 $ & $ 0.67\pm 0.02   $ & $ -417.1\pm 0.4 $ & $ 18.9\pm 0.5 $ & 2005sep \\
M31         & M2357      & $   5.8\pm 0.2 $ & $ 0.58\pm 0.02   $ & $ -561.7\pm 0.5 $ & $ 16.1\pm 0.7 $ & 2004nov \\
M31         & M2371      & $   8.6\pm 0.2 $ & $ 0.67\pm 0.02   $ & $ -463.5\pm 0.4 $ & $ 19.0\pm 0.5 $ & 2004nov \\
M31         & M2401      & $   9.1\pm 0.2 $ & $ 0.56\pm 0.01   $ & $ -437.0\pm 0.3 $ & $ 15.5\pm 0.3 $ & 2004nov \\
M31         & M2404      & $  21.0\pm 0.3 $ & $ 0.65\pm 0.01   $ & $ -379.6\pm 0.3 $ & $ 18.2\pm 0.4 $ & 2004nov \\
M31         & M2410      & $  13.1\pm 0.5 $ & $ 0.92\pm 0.04   $ & $ -436.2\pm 0.9 $ & $ 27\pm 1     $ & 2004nov \\
M31         & M2437      & $  12.2\pm 0.3 $ & $ 0.74\pm 0.02   $ & $  -73.8\pm 0.5 $ & $ 21.2\pm 0.6 $ & 2004nov \\
M31         & M2466      & $   4.8\pm 0.2 $ & $ 0.46\pm 0.02   $ & $ -412.5\pm 0.5 $ & $ 12.2\pm 0.6 $ & 2005sep \\
M31         & M2496      & $  11.7\pm 0.2 $ & $ 0.534\pm 0.009 $ & $ -353.7\pm 0.2 $ & $ 14.6\pm 0.3 $ & 2004nov \\
M31         & M2501      & $   3.2\pm 0.2 $ & $ 0.66\pm 0.06   $ & $ -386\pm 1     $ & $ 19\pm 1     $ & 2004nov \\
M31         & M2502      & $   3.6\pm 0.2 $ & $ 0.65\pm 0.04   $ & $ -428.5\pm 0.9 $ & $ 19\pm 1     $ & 2004nov \\
M31         & M2507      & $   9.8\pm 0.3 $ & $ 0.61\pm 0.02   $ & $ -184.7\pm 0.4 $ & $ 16.9\pm 0.5 $ & 2004nov \\
M31         & M2512      & $   6.6\pm 0.3 $ & $ 0.61\pm 0.03   $ & $ -315.6\pm 0.8 $ & $ 17\pm 1     $ & 2004nov \\
M31         & M2514      & $   4.1\pm 0.3 $ & $ 0.61\pm 0.04   $ & $ -453\pm 1     $ & $ 17\pm 1     $ & 2004nov \\
M31         & M2519      & $   6.0\pm 0.2 $ & $ 0.55\pm 0.02   $ & $ -480.4\pm 0.4 $ & $ 15.2\pm 0.5 $ & 2004nov \\
M31         & M2538      & $  11.5\pm 0.1 $ & $ 0.434\pm 0.006 $ & $ -444.3\pm 0.1 $ & $ 11.3\pm 0.2 $ & 2007aug \\
M31         & M2694      & $  10.2\pm 0.2 $ & $ 0.53\pm 0.01   $ & $ -240.5\pm 0.3 $ & $ 14.6\pm 0.4 $ & 2005sep \\
M31         & M2860      & $   9.2\pm 0.2 $ & $ 0.68\pm 0.02   $ & $ -413.4\pm 0.4 $ & $ 19.3\pm 0.5 $ & 2005sep \\
M31         & M2943      & $   8.9\pm 0.2 $ & $ 0.53\pm 0.01   $ & $ -329.4\pm 0.3 $ & $ 14.6\pm 0.3 $ & 2007aug \\
M31         & M2985      & $   6.7\pm 0.2 $ & $ 0.46\pm 0.01   $ & $ -431.9\pm 0.3 $ & $ 12.2\pm 0.4 $ & 2004nov \\
M31         & M2988      & $   5.0\pm 0.2 $ & $ 0.49\pm 0.02   $ & $ -116.7\pm 0.6 $ & $ 13.0\pm 0.7 $ & 2004nov \\
M31         & M3246      & $   7.8\pm 0.2 $ & $ 0.68\pm 0.02   $ & $ -542.9\pm 0.4 $ & $ 19.3\pm 0.5 $ & 2005sep \\
M31         & PN001      & $   7.0\pm 0.3 $ & $ 0.66\pm 0.03   $ & $ -418.8\pm 0.8 $ & $ 19.0\pm 1   $ & 2001sep \\
M31         & PN003      & $   4.9\pm 0.3 $ & $ 0.56\pm 0.04   $ & $ -273\pm 1     $ & $ 15\pm 1     $ & 2006sep \\
M31         & PN008      & $   5.7\pm 0.3 $ & $ 0.65\pm 0.04   $ & $ -241\pm 1     $ & $ 18\pm 1     $ & 2006sep \\
M31         & PN010      & $   4.5\pm 0.3 $ & $ 0.61\pm 0.05   $ & $ -157\pm 1     $ & $ 17\pm 2     $ & 2001sep \\
M31         & PN012      & $   4.7\pm 0.3 $ & $ 0.79\pm 0.06   $ & $ -236\pm 2     $ & $ 23\pm 2     $ & 2007jan \\
M31         & PN015      & $  10.5\pm 0.6 $ & $ 0.99\pm 0.06   $ & $ -351\pm 2     $ & $ 29\pm 2     $ & 2001sep, 2005sep \\
M31         & PN017      & $   2.1\pm 0.2 $ & $ 0.61\pm 0.05   $ & $ -495\pm 1     $ & $ 17\pm 1     $ & 2001sep \\
M31         & PN018      & $   3.9\pm 0.3 $ & $ 0.62\pm 0.05   $ & $ -507\pm 1     $ & $ 17\pm 1     $ & 2001sep \\
M31         & PN023      & $   7.1\pm 0.3 $ & $ 0.72\pm 0.03   $ & $ -231.2\pm 0.8 $ & $ 20\pm 1     $ & 2006sep \\
M31         & PN024      & $   4.9\pm 0.2 $ & $ 0.69\pm 0.04   $ & $ -273.5\pm 0.9 $ & $ 20\pm 1     $ & 2001sep, 2005sep \\
M31         & PN026      & $   6.3\pm 0.3 $ & $ 0.56\pm 0.03   $ & $ -400.2\pm 0.7 $ & $ 15.4\pm 0.9 $ & 2001sep \\
M31         & PN027      & $   6.4\pm 0.2 $ & $ 0.89\pm 0.03   $ & $ -135.3\pm 0.8 $ & $ 26\pm 1     $ & 2006sep \\
M31         & PN028      & $   4.5\pm 0.3 $ & $ 0.80\pm 0.07   $ & $ -256\pm 2     $ & $ 23\pm 2     $ & 2001sep \\
M31         & PN029      & $   5.8\pm 0.4 $ & $ 0.97\pm 0.06   $ & $ -456\pm 2     $ & $ 28\pm 2     $ & 2001sep, 2005sep \\
M31         & PN030      & $  11.4\pm 0.4 $ & $ 1.03\pm 0.04   $ & $ -279\pm 1     $ & $ 30\pm 1     $ & 2001sep, 2007jan \\
M31         & PN031      & $   6.2\pm 0.2 $ & $ 0.57\pm 0.02   $ & $ -226.5\pm 0.6 $ & $ 15.8\pm 0.7 $ & 2001sep \\
M31         & PN032      & $  16.8\pm 0.6 $ & $ 0.90\pm 0.03   $ & $ -119.5\pm 0.8 $ & $ 26\pm 1     $ & 2001sep, 2005sep \\
M31         & PN033      & $   4.8\pm 0.2 $ & $ 0.63\pm 0.03   $ & $  131.7\pm 0.8 $ & $ 18\pm 1     $ & 2007jan \\
M31         & PN035      & $   4.4\pm 0.3 $ & $ 0.71\pm 0.05   $ & $ -372\pm 1     $ & $ 20\pm 1     $ & 2001sep, 2005sep \\
M31         & PN036      & $   1.9\pm 0.2 $ & $ 0.58\pm 0.07   $ & $ -183\pm 2     $ & $ 16\pm 2     $ & 2001sep \\
M31         & PN037      & $   1.9\pm 0.3 $ & $ 0.8\pm 0.1     $ & $ -346\pm 3     $ & $ 23\pm 4     $ & 2001sep \\
M31         & PN038      & $   5.2\pm 0.2 $ & $ 0.71\pm 0.03   $ & $ -319.1\pm 0.8 $ & $ 20\pm 1     $ & 2007jan \\
M31         & PN042      & $   8.7\pm 0.2 $ & $ 0.55\pm 0.01   $ & $ -390.1\pm 0.3 $ & $ 15.2\pm 0.3 $ & 2005jul \\
M31         & PN043      & $   5.8\pm 0.2 $ & $ 0.63\pm 0.02   $ & $  -93.3\pm 0.5 $ & $ 17.7\pm 0.7 $ & 2006sep \\
M31         & PN045      & $  14.8\pm 0.3 $ & $ 0.55\pm 0.01   $ & $ -255.0\pm 0.3 $ & $ 15.1\pm 0.4 $ & 2001sep \\
M31         & PN046      & $  14.6\pm 0.4 $ & $ 0.57\pm 0.02   $ & $ -457.2\pm 0.4 $ & $ 15.7\pm 0.5 $ & 2005jul \\
M31         & PN047      & $  14.2\pm 0.4 $ & $ 0.61\pm 0.02   $ & $ -316.7\pm 0.5 $ & $ 17.2\pm 0.6 $ & 2005jul \\
M31         & PN048      & $  14.6\pm 0.2 $ & $ 0.76\pm 0.01   $ & $ -237.5\pm 0.3 $ & $ 21.8\pm 0.3 $ & 2005sep \\
M31         & PN049      & $   3.2\pm 0.2 $ & $ 0.59\pm 0.04   $ & $ -334\pm 1     $ & $ 17\pm 1     $ & 2005jul \\
M31         & PN052      & $   4.8\pm 0.2 $ & $ 0.59\pm 0.03   $ & $ -153.4\pm 0.6 $ & $ 16.5\pm 0.8 $ & 2005sep \\
M31         & PN053      & $   7.1\pm 0.2 $ & $ 0.45\pm 0.01   $ & $ -270.3\pm 0.4 $ & $ 11.8\pm 0.4 $ & 2001sep \\
M31         & PN054      & $   8.3\pm 0.2 $ & $ 0.48\pm 0.01   $ & $ -499.7\pm 0.3 $ & $ 13.0\pm 0.4 $ & 2005sep \\
M31         & PN055      & $   4.1\pm 0.2 $ & $ 0.75\pm 0.04   $ & $ -492.8\pm 0.9 $ & $ 22\pm 1     $ & 2005sep \\
M31         & PN056      & $   7.6\pm 0.2 $ & $ 0.54\pm 0.01   $ & $ -252.1\pm 0.4 $ & $ 14.7\pm 0.4 $ & 2005sep \\
M31         & PN058      & $   5.5\pm 0.3 $ & $ 0.74\pm 0.04   $ & $ -420\pm 1     $ & $ 21\pm 1     $ & 2006sep \\
M31         & PN061      & $   8.7\pm 0.3 $ & $ 0.67\pm 0.02   $ & $ -183.4\pm 0.6 $ & $ 19.0\pm 0.7 $ & 2001sep, 2005sep \\
M31         & PN062      & $   8.3\pm 0.3 $ & $ 0.69\pm 0.02   $ & $ -570.2\pm 0.6 $ & $ 19.6\pm 0.7 $ & 2006sep \\
M31         & PN064      & $   2.3\pm 0.2 $ & $ 0.61\pm 0.07   $ & $ -609\pm   2   $ & $ 17\pm 2     $ & 2001sep \\
M31         & PN067      & $  19.0\pm 0.3 $ & $ 0.498\pm 0.008 $ & $ -410.6\pm 0.2 $ & $ 13.4\pm 0.2 $ & 2005sep \\
M31         & PN069      & $   3.8\pm 0.2 $ & $ 0.76\pm 0.05   $ & $ -522\pm 1     $ & $ 22\pm 2     $ & 2005sep \\
M31         & PN071      & $   2.5\pm 0.2 $ & $ 0.71\pm 0.05   $ & $ -229\pm 1     $ & $ 20\pm 2     $ & 2007jan \\
M31         & PN072      & $   5.4\pm 0.4 $ & $ 0.54\pm 0.04   $ & $ -351\pm 1     $ & $ 15\pm 1     $ & 2005sep \\
M31         & PN075      & $   5.6\pm 0.4 $ & $ 0.81\pm 0.06   $ & $ -398\pm 1     $ & $ 23\pm 2     $ & 2007jan \\
M31         & PN080      & $  11.0\pm 0.3 $ & $ 0.73\pm 0.02   $ & $ -299.7\pm 0.5 $ & $ 20.9\pm 0.6 $ & 2007jan \\
M31         & PN087      & $   3.4\pm 0.3 $ & $ 0.70\pm 0.07   $ & $ -381\pm 2     $ & $ 20\pm 2     $ & 2001sep, 2007jan \\
M31         & PN091      & $   6.5\pm 0.2 $ & $ 0.80\pm 0.03   $ & $ -182.0\pm 0.7 $ & $ 23.0\pm 0.8 $ & 2007jan \\
M31         & PN092      & $   8.0\pm 0.2 $ & $ 0.49\pm 0.01   $ & $ -275.0\pm 0.3 $ & $ 13.1\pm 0.3 $ & 2005sep \\
M31         & PN093      & $   4.5\pm 0.2 $ & $ 0.85\pm 0.05   $ & $ -546\pm 1     $ & $ 25\pm 1     $ & 2006sep \\
M31         & PN095      & $   4.4\pm 0.1 $ & $ 0.42\pm 0.02   $ & $ -394.3\pm 0.4 $ & $ 10.8\pm 0.4 $ & 2006sep \\
M31         & PN097      & $   8.4\pm 0.2 $ & $ 0.64\pm 0.01   $ & $ -480.0\pm 0.3 $ & $ 17.9\pm 0.4 $ & 2005sep \\
M31         & PN116      & $   8.9\pm 0.2 $ & $ 0.66\pm 0.02   $ & $ -350.1\pm 0.4 $ & $ 18.8\pm 0.5 $ & 2005sep \\
M31         & PN125      & $   5.0\pm 0.2 $ & $ 0.55\pm 0.03   $ & $ -227.6\pm 0.7 $ & $ 15.1\pm 0.8 $ & 2006sep \\
M31         & PN131      & $   5.0\pm 0.1 $ & $ 0.53\pm 0.02   $ & $ -339.3\pm 0.4 $ & $ 14.4\pm 0.5 $ & 2005sep \\
M31         & pn136      & $   3.4\pm 0.3 $ & $ 0.73\pm 0.06   $ & $ -329\pm 2     $ & $ 21\pm 2     $ & 2007jan \\
M31         & PN142      & $   3.1\pm 0.2 $ & $ 0.62\pm 0.04   $ & $ -259\pm 1     $ & $ 17\pm 1     $ & 2005jul \\
M31         & PN143      & $   1.6\pm 0.2 $ & $ 1.0\pm 0.1     $ & $ -267\pm 3     $ & $ 28\pm 4     $ & 2005sep \\
M31         & PN150      & $   5.4\pm 0.2 $ & $ 0.64\pm 0.02   $ & $ -136.2\pm 0.6 $ & $ 18.2\pm 0.7 $ & 2005sep \\
M31         & PN151      & $   6.8\pm 0.2 $ & $ 0.76\pm 0.02   $ & $ -562.1\pm 0.6 $ & $ 22.0\pm 0.7 $ & 2005sep \\
M31         & PN154      & $   7.3\pm 0.2 $ & $ 0.55\pm 0.01   $ & $ -349.3\pm 0.3 $ & $ 15.2\pm 0.4 $ & 2005sep \\
M31         & PN155      & $   9.8\pm 0.2 $ & $ 0.69\pm 0.01   $ & $ -241.6\pm 0.3 $ & $ 19.5\pm 0.4 $ & 2005sep \\
M31         & PN172      & $   8.9\pm 0.2 $ & $ 0.82\pm 0.02   $ & $ -247.2\pm 0.5 $ & $ 23.6\pm 0.6 $ & 2005sep \\
M31         & PN177      & $   4.1\pm 0.2 $ & $ 0.58\pm 0.04   $ & $ -300.3\pm 0.9 $ & $ 16\pm 1     $ & 2007jan \\
M31         & PN178      & $  11.8\pm 0.2 $ & $ 0.58\pm 0.01   $ & $ -213.8\pm 0.3 $ & $ 16.1\pm 0.3 $ & 2005sep \\
M31         & pn179      & $   4.0\pm 0.2 $ & $ 0.62\pm 0.04   $ & $ -124.6\pm 0.9 $ & $ 17\pm 1     $ & 2007jan \\
M31         & PN190      & $   4.8\pm 0.2 $ & $ 0.48\pm 0.02   $ & $ -113.2\pm 0.5 $ & $ 12.9\pm 0.7 $ & 2007jan \\
M31         & PN209      & $   6.1\pm 0.2 $ & $ 0.47\pm 0.01   $ & $ -320.9\pm 0.3 $ & $ 12.6\pm 0.4 $ & 2005sep \\
M31         & PN216      & $   7.2\pm 0.2 $ & $ 0.43\pm 0.01   $ & $ -229.1\pm 0.3 $ & $ 11.2\pm 0.3 $ & 2005sep \\
M31         & PN219      & $   5.1\pm 0.2 $ & $ 0.71\pm 0.03   $ & $ -140.9\pm 0.7 $ & $ 20.2\pm 0.8 $ & 2005sep \\
M31         & PN237      & $   5.3\pm 0.2 $ & $ 0.87\pm 0.04   $ & $  -64.7\pm 0.9 $ & $ 25\pm 1     $ & 2005sep \\
M31         & PN240      & $  11.0\pm 0.2 $ & $ 0.61\pm 0.01   $ & $  -95.1\pm 0.3 $ & $ 17.2\pm 0.4 $ & 2005sep \\
M31         & PN278      & $   4.2\pm 0.3 $ & $ 0.75\pm 0.05   $ & $ -135\pm 1     $ & $ 21\pm 2     $ & 2005sep \\
M31         & PN290      & $   3.4\pm 0.2 $ & $ 0.57\pm 0.03   $ & $ -658.6\pm 0.7 $ & $ 15.7\pm 0.9 $ & 2007jan \\
M31         & PN335      & $   3.5\pm 0.2 $ & $ 0.74\pm 0.04   $ & $ -446.1\pm 0.9 $ & $ 21\pm 1     $ & 2004nov \\
M31         & PN344      & $   8.5\pm 0.2 $ & $ 0.60\pm 0.01   $ & $ -505.2\pm 0.3 $ & $ 16.7\pm 0.4 $ & 2004nov \\
M31         & PN345      & $   9.8\pm 0.2 $ & $ 0.60\pm 0.02   $ & $ -490.4\pm 0.4 $ & $ 16.8\pm 0.5 $ & 2004nov \\
M31         & PN349      & $   6.4\pm 0.2 $ & $ 0.47\pm 0.01   $ & $ -494.5\pm 0.3 $ & $ 12.6\pm 0.4 $ & 2004nov \\
M31         & PN353      & $   6.9\pm 0.2 $ & $ 0.74\pm 0.02   $ & $ -257.3\pm 0.5 $ & $ 21.1\pm 0.6 $ & 2004nov \\
M31         & PN363      & $   7.2\pm 0.2 $ & $ 0.75\pm 0.02   $ & $  -95.6\pm 0.5 $ & $ 21.6\pm 0.6 $ & 2007aug \\
M31         & PN364      & $   6.8\pm 0.1 $ & $ 0.309\pm 0.006 $ & $ -102.5\pm 0.1 $ & $  6.6\pm 0.2 $ & 2005sep \\
M31         & PN370      & $   5.7\pm 0.2 $ & $ 0.41\pm 0.01   $ & $  -91.0\pm 0.3 $ & $ 10.4\pm 0.4 $ & 2004nov \\
M31         & PN375      & $  11.5\pm 0.2 $ & $ 0.71\pm 0.02   $ & $ -410.6\pm 0.4 $ & $ 20.1\pm 0.4 $ & 2005sep \\
M31         & PN380      & $  13.3\pm 0.2 $ & $ 0.66\pm 0.01   $ & $ -297.1\pm 0.3 $ & $ 18.7\pm 0.3 $ & 2005sep \\
M31         & PN387      & $   7.5\pm 0.2 $ & $ 0.66\pm 0.02   $ & $ -254.6\pm 0.4 $ & $ 18.6\pm 0.5 $ & 2005sep \\
M31         & PN390      & $   9.8\pm 0.2 $ & $ 0.83\pm 0.02   $ & $ -578.7\pm 0.5 $ & $ 23.9\pm 0.6 $ & 2005sep \\
M31         & PN410      & $   6.1\pm 0.2 $ & $ 0.82\pm 0.03   $ & $ -530.3\pm 0.7 $ & $ 23.7\pm 0.8 $ & 2005sep \\
M31         & PN413      & $  10.5\pm 0.3 $ & $ 0.52\pm 0.01   $ & $ -317.6\pm 0.4 $ & $ 14.2\pm 0.4 $ & 2005sep \\
M31         & PN414      & $   5.1\pm 0.2 $ & $ 0.73\pm 0.03   $ & $   37.6\pm 0.8 $ & $ 20.7\pm 0.9 $ & 2005sep \\
M31         & PN450      & $   2.3\pm 0.2 $ & $ 0.86\pm 0.06   $ & $ -361\pm 1     $ & $ 25\pm 2     $ & 2005sep \\
M31         & PN478      & $   5.6\pm 0.2 $ & $ 0.47\pm 0.02   $ & $ -115.9\pm 0.5 $ & $ 12.6\pm 0.6 $ & 2007jan \\
M31         & PN537      & $   7.8\pm 0.1 $ & $ 0.425\pm 0.009 $ & $ -154.8\pm 0.2 $ & $ 11.0\pm 0.3 $ & 2005sep \\
M31         & PN555      & $  11.2\pm 0.2 $ & $ 0.80\pm 0.02   $ & $  -98.5\pm 0.4 $ & $ 23.1\pm 0.6 $ & 2005sep \\
M31         & PN557      & $   7.1\pm 0.2 $ & $ 0.66\pm 0.02   $ & $  -70.2\pm 0.5 $ & $ 18.8\pm 0.6 $ & 2005sep \\
M31         & PN559      & $  11.9\pm 0.1 $ & $ 0.520\pm 0.007 $ & $ -113.0\pm 0.2 $ & $ 14.2\pm 0.2 $ & 2005sep \\
M31         & PN563      & $   8.8\pm 0.3 $ & $ 0.87\pm 0.03   $ & $ -157.5\pm 0.7 $ & $ 25.1\pm 0.9 $ & 2005sep \\
M31         & PN568      & $   7.9\pm 0.2 $ & $ 0.69\pm 0.02   $ & $ -118.7\pm 0.5 $ & $ 19.8\pm 0.7 $ & 2005sep \\
M31         & PN569      & $   2.7\pm 0.2 $ & $ 0.87\pm 0.06   $ & $ -148\pm 2     $ & $ 25\pm 2     $ & 2005sep \\
M32         & PN1        & $  10.6\pm 0.3 $ & $ 0.56\pm 0.02   $ & $ -206.7\pm 0.4 $ & $ 15.3\pm 0.5 $ & 2001sep, 2006sep \\
M32         & PN2        & $   6.6\pm 0.2 $ & $ 0.66\pm 0.02   $ & $ -218.0\pm 0.6 $ & $ 18.8\pm 0.7 $ & 2006sep \\
M32         & PN3        & $   5.6\pm 0.2 $ & $ 0.46\pm 0.02   $ & $ -169.3\pm 0.5 $ & $ 12.3\pm 0.6 $ & 2001sep \\
M32         & PN5        & $   5.5\pm 0.4 $ & $ 0.80\pm 0.06   $ & $ -245\pm 1     $ & $ 23\pm 2     $ & 2001sep, 2006sep \\
M32         & PN6        & $   4.8\pm 0.3 $ & $ 0.84\pm 0.06   $ & $ -206\pm 1     $ & $ 24\pm 2     $ & 2001sep, 2006sep \\
M32         & PN7        & $   5.6\pm 0.2 $ & $ 0.42\pm 0.02   $ & $ -194.4\pm 0.5 $ & $ 10.8\pm 0.6 $ & 2001sep, 2006sep \\
M32         & PN8        & $   4.0\pm 0.3 $ & $ 0.83\pm 0.06   $ & $ -168\pm 2     $ & $ 24\pm 2     $ & 2001sep \\
M32         & PN21       & $  10.1\pm 0.5 $ & $ 0.90\pm 0.05   $ & $ -192\pm 1     $ & $ 26\pm 1     $ & 2001sep, 2006sep \\
M32         & PN23       & $   3.5\pm 0.2 $ & $ 0.87\pm 0.05   $ & $ -231\pm 1     $ & $ 25\pm 2     $ & 2006sep \\
M32         & PN24+PN25  & $   7.8\pm 0.2 $ & $ 0.77\pm 0.03   $ & $ -171.1\pm 0.6 $ & $ 22.2\pm 0.7 $ & 2006sep \\
M32         & PN24       & $   3.8\pm 0.2 $ & $ 0.82\pm 0.04   $ & $ -166.8\pm 0.9 $ & $ 24\pm 1     $ & 2006sep \\
M32         & PN25       & $   4.1\pm 0.2 $ & $ 0.72\pm 0.03   $ & $ -174.8\pm 0.7 $ & $ 20.5\pm 0.9 $ & 2006sep \\
M32         & PN26       & $   3.5\pm 0.2 $ & $ 0.91\pm 0.06   $ & $ -229\pm 2     $ & $ 26\pm 2     $ & 2006sep \\
M32         & PN29       & $   2.4\pm 0.2 $ & $ 0.63\pm 0.06   $ & $ -646\pm 2     $ & $ 18\pm 2     $ & 2006sep \\
M33         & M002       & $   5.0\pm 0.3 $ & $ 0.56\pm 0.03   $ & $ -126.9\pm 0.8 $ & $ 16\pm 1     $ & 2003oct \\
M33         & M005       & $   4.0\pm 0.2 $ & $ 0.38\pm 0.02   $ & $ -125.5\pm 0.6 $ & $  9.5\pm 0.7 $ & 2003oct \\
M33         & M008       & $   4.4\pm 0.2 $ & $ 0.45\pm 0.02   $ & $ -115.7\pm 0.5 $ & $ 11.9\pm 0.6 $ & 2003oct \\
M33         & M013       & $   3.2\pm 0.1 $ & $ 0.44\pm 0.02   $ & $ -106.9\pm 0.6 $ & $ 11.5\pm 0.7 $ & 2002nov \\
M33         & M017       & $   7.0\pm 0.3 $ & $ 0.57\pm 0.03   $ & $ -195.9\pm 0.7 $ & $ 15.8\pm 0.8 $ & 2002nov \\
M33         & M018       & $  13.4\pm 0.3 $ & $ 0.55\pm 0.01   $ & $ -155.2\pm 0.3 $ & $ 15.3\pm 0.4 $ & 2002nov \\
M33         & M028       & $   5.0\pm 0.2 $ & $ 0.54\pm 0.03   $ & $ -161.4\pm 0.7 $ & $ 15.0\pm 0.9 $ & 2002nov \\
M33         & M042       & $  14.4\pm 0.5 $ & $ 0.79\pm 0.03   $ & $ -136.1\pm 0.7 $ & $ 22.7\pm 0.8 $ & 2002nov \\
M33         & M046       & $   5.0\pm 0.2 $ & $ 0.37\pm 0.02   $ & $ -127.7\pm 0.4 $ & $  8.9\pm 0.5 $ & 2002nov \\
M33         & M059       & $   5.1\pm 0.2 $ & $ 0.40\pm 0.02   $ & $ -113.3\pm 0.4 $ & $ 10.0\pm 0.5 $ & 2003oct \\
M33         & M061       & $   6.8\pm 0.3 $ & $ 0.66\pm 0.03   $ & $ -115.3\pm 0.8 $ & $ 18.7\pm 0.9 $ & 2002nov \\
M33         & M062       & $   4.4\pm 0.2 $ & $ 0.53\pm 0.03   $ & $  -98.3\pm 0.8 $ & $ 15\pm 1     $ & 2002nov \\
M33         & M063       & $   8.2\pm 0.3 $ & $ 0.68\pm 0.03   $ & $ -111.5\pm 0.7 $ & $ 19.2\pm 0.8 $ & 2002nov \\
M33         & M065       & $   4.9\pm 0.2 $ & $ 0.54\pm 0.03   $ & $ -170.7\pm 0.7 $ & $ 14.7\pm 0.9 $ & 2002nov \\
M33         & M067       & $  11.0\pm 0.3 $ & $ 0.64\pm 0.02   $ & $ -128.0\pm 0.5 $ & $ 18.1\pm 0.6 $ & 2002nov \\
M33         & M068       & $   7.1\pm 0.3 $ & $ 0.50\pm 0.03   $ & $ -170.7\pm 0.6 $ & $ 13.4\pm 0.8 $ & 2002nov \\
M33         & M069       & $  10.5\pm 0.5 $ & $ 0.84\pm 0.05   $ & $ -168\pm 1     $ & $ 24\pm 1     $ & 2002nov \\
M33         & M072       & $   3.8\pm 0.2 $ & $ 0.53\pm 0.03   $ & $ -267.5\pm 0.8 $ & $ 14.5\pm 0.9 $ & 2002nov \\
M33         & M074       & $   4.0\pm 0.3 $ & $ 0.72\pm 0.06   $ & $ -168\pm 2     $ & $ 21\pm 2     $ & 2003oct \\
M33         & M075       & $   5.8\pm 0.2 $ & $ 0.46\pm 0.02   $ & $ -259.8\pm 0.4 $ & $ 12.3\pm 0.5 $ & 2002nov \\
M33         & M079       & $   6.1\pm 0.3 $ & $ 0.75\pm 0.04   $ & $ -165\pm 1     $ & $ 21\pm 1     $ & 2003oct \\
M33         & M089       & $   4.1\pm 0.3 $ & $ 0.69\pm 0.05   $ & $ -272\pm 1     $ & $ 20\pm 2     $ & 2003oct \\
M33         & M091       & $  24.0\pm 0.5 $ & $ 0.70\pm 0.01   $ & $ -137.6\pm 0.4 $ & $ 19.8\pm 0.4 $ & 2002nov, 2003oct \\
M33         & M093       & $  14.2\pm 0.4 $ & $ 0.66\pm 0.02   $ & $ -171.6\pm 0.6 $ & $ 18.8\pm 0.7 $ & 2002nov, 2003oct \\
M33         & M094       & $  13.4\pm 0.4 $ & $ 0.57\pm 0.02   $ & $ -168.3\pm 0.5 $ & $ 15.8\pm 0.5 $ & 2002nov, 2003oct \\
M33         & M095       & $   4.1\pm 0.2 $ & $ 0.46\pm 0.03   $ & $ -262.5\pm 0.7 $ & $ 12.1\pm 0.8 $ & 2002nov \\
M33         & M096       & $   8.6\pm 0.2 $ & $ 0.44\pm 0.01   $ & $ -151.7\pm 0.3 $ & $ 11.5\pm 0.4 $ & 2003oct \\
M33         & M101       & $   7.1\pm 0.2 $ & $ 0.51\pm 0.02   $ & $ -132.7\pm 0.4 $ & $ 13.9\pm 0.5 $ & 2003oct \\
M33         & M111       & $   8.9\pm 0.3 $ & $ 0.70\pm 0.03   $ & $ -241.3\pm 0.7 $ & $ 20.0\pm 0.8 $ & 2003oct \\
M33         & M119       & $   8.1\pm 0.3 $ & $ 0.81\pm 0.04   $ & $ -231.4\pm 0.9 $ & $ 23\pm 1     $ & 2003oct \\
M33         & M125       & $   3.3\pm 0.2 $ & $ 0.64\pm 0.04   $ & $ -182\pm 1     $ & $ 18\pm 1     $ & 2003oct \\
M33         & M128       & $   4.8\pm 0.2 $ & $ 0.34\pm 0.01   $ & $ -260.1\pm 0.3 $ & $  8.0\pm 0.4 $ & 2002nov \\
M33         & newPN      & $   6.2\pm 0.2 $ & $ 0.35\pm 0.01   $ & $ -112.5\pm 0.2 $ & $  8.1\pm 0.3 $ & 2003oct \\
NGC147      & PN04       & $   2.8\pm 0.1 $ & $ 0.34\pm 0.02   $ & $ -192.2\pm 0.5 $ & $  7.7\pm 0.6 $ & 2007jan \\
NGC147      & PN07       & $   3.8\pm 0.2 $ & $ 0.53\pm 0.03   $ & $ -182.5\pm 0.9 $ & $ 15\pm 1     $ & 2007jan \\
NGC185      & PN01       & $  10.5\pm 0.3 $ & $ 0.73\pm 0.02   $ & $ -232.7\pm 0.6 $ & $ 20.9\pm 0.7 $ & 2001sep \\
NGC185      & PN02       & $   4.3\pm 0.1 $ & $ 0.37\pm 0.01   $ & $ -206.0\pm 0.4 $ & $  8.8\pm 0.4 $ & 2001sep \\
NGC185      & PN03       & $   6.2\pm 0.3 $ & $ 0.60\pm 0.03   $ & $ -213.1\pm 0.7 $ & $ 16.8\pm 0.8 $ & 2001sep, 2006sep \\
NGC185      & PN04       & $   3.5\pm 0.3 $ & $ 0.71\pm 0.07   $ & $ -215\pm 2     $ & $ 20\pm 2     $ & 2006sep \\
NGC185      & PN05       & $   7.1\pm 0.4 $ & $ 0.88\pm 0.05   $ & $ -245\pm 1     $ & $ 26\pm 1     $ & 2001sep, 2006sep \\
NGC205      & PN1        & $   4.5\pm 0.2 $ & $ 0.31\pm 0.02   $ & $ -291.2\pm 0.4 $ & $  6.7\pm 0.4 $ & 2001sep \\
NGC205      & PN2        & $   5.2\pm 0.3 $ & $ 0.74\pm 0.04   $ & $ -230\pm 1     $ & $ 21\pm 1     $ & 2007aug \\
NGC205      & PN4        & $   4.7\pm 0.2 $ & $ 0.47\pm 0.03   $ & $ -251.1\pm 0.6 $ & $ 12.6\pm 0.7 $ & 2001sep \\
NGC205      & PN5        & $   6.7\pm 0.3 $ & $ 0.58\pm 0.03   $ & $ -240.6\pm 0.6 $ & $ 16.0\pm 0.8 $ & 2001sep \\
NGC205      & PN6        & $   4.8\pm 0.3 $ & $ 0.57\pm 0.04   $ & $ -210.1\pm 0.9 $ & $ 16\pm 1     $ & 2001sep \\
NGC205      & PN7        & $   5.2\pm 0.2 $ & $ 0.33\pm 0.01   $ & $ -243.7\pm 0.3 $ & $  7.6\pm 0.3 $ & 2001sep \\
NGC205      & PN8        & $   3.7\pm 0.2 $ & $ 0.58\pm 0.04   $ & $ -259.2\pm 0.9 $ & $ 16\pm 1     $ & 2001sep \\
NGC205      & PN9        & $   3.6\pm 0.2 $ & $ 0.36\pm 0.02   $ & $ -241.4\pm 0.4 $ & $  8.5\pm 0.5 $ & 2001sep \\
NGC205      & PN10       & $   5.4\pm 0.3 $ & $ 0.79\pm 0.04   $ & $ -220\pm 1     $ & $ 23\pm 1     $ & 2007aug \\
NGC6822     & PN1        & $   4.5\pm 0.2 $ & $ 0.48\pm 0.02   $ & $  -73.0\pm 0.4 $ & $ 12.9\pm 0.5 $ & 2005jul \\
NGC6822     & PN19       & $   4.4\pm 0.2 $ & $ 0.48\pm 0.03   $ & $  -69.0\pm 0.7 $ & $ 12.8\pm 0.8 $ & 2005jul \\
NGC6822     & S14        & $   5.0\pm 0.3 $ & $ 0.62\pm 0.04   $ & $ -100.7\pm 0.9 $ & $ 17\pm 1     $ & 2002jul \\
NGC6822     & S16        & $  11.3\pm 0.5 $ & $ 0.60\pm 0.03   $ & $  -78.7\pm 0.7 $ & $ 16.7\pm 0.8 $ & 2002jul \\
NGC6822     & S30        & $   4.2\pm 0.2 $ & $ 0.97\pm 0.06   $ & $  -60\pm 1     $ & $ 28\pm 2     $ & 2005jul \\
NGC6822     & S33        & $   7.5\pm 0.4 $ & $ 0.85\pm 0.05   $ & $  -43\pm 1     $ & $ 25\pm 2     $ & 2002jul \\
Sagittarius & He 2-436   & $1480\pm 20    $ & $ 0.474\pm 0.006 $ & $  131.6\pm 0.1 $ & $ 12.6\pm 0.2 $ & 2004juna \\
Sagittarius & StWr 2-21  & $ 218\pm 3     $ & $ 0.94\pm 0.01   $ & $  116.7\pm 0.3 $ & $ 27.2\pm 0.4 $ & 2004junc \\
Sagittarius & Wray 16-423 & $3200\pm 30   $ & $ 0.727\pm 0.007 $ & $  131.6\pm 0.2 $ & $ 20.8\pm 0.2 $ & 2004juna \\
Sextans A   & PN1        & $   5.0\pm 0.3 $ & $ 0.88\pm 0.06   $ & $  315\pm 1     $ & $ 26\pm 2     $ & 2004nov \\
\end{longtable}

\label{tab_o3}
\clearpage

\begin{table*}[!t]\centering
  \setlength{\tabnotewidth}{0.9\columnwidth}
  \tablecols{7}
  \caption{H$\alpha$ data for extragalactic planetary nebulae} \label{tab_Ha}
 \begin{tabular}{llccccl}
    \toprule
Galaxy  & Object & Flux\ $^{\mathrm a}$   & FWHM  & $V_{helio}\ ^{\mathrm b}$ & $\Delta V_{0.5}$ & Run     \\
        &        & $(10^3\,\mathrm{ADU})$ & (\AA) & (km/s)      & (km/s)           &         \\
    \midrule
Fornax      & PN1        & $  35.4\pm 0.2 $ & $ 0.723\pm 0.005 $ & $   47.95\pm 0.09 $ & $ 11.4\pm 0.1   $ & 2004nov  \\
M32         & PN3        & $   4.0\pm 0.4 $ & $ 0.9\pm 0.1     $ & $ -157\pm 2       $ & $ 17\pm 2       $ & 2001sep  \\
M32         & PN6        & $   4.3\pm 0.5 $ & $ 1.5\pm 0.2     $ & $ -185\pm 3       $ & $ 33\pm 4       $ & 2001sep  \\
M33         & M059       & $   2.2\pm 0.3 $ & $ 0.8\pm 0.1     $ & $ -119\pm 2       $ & $ 13\pm 2       $ & 2003oct  \\
NGC6822     & PN19       & $   4.8\pm 0.2 $ & $ 0.82\pm 0.03   $ & $  -70.5\pm 0.5   $ & $ 14.3\pm 0.7   $ & 2005jul  \\
NGC6822     & S14        & $   3.1\pm 0.3 $ & $ 0.97\pm 0.09   $ & $ -104\pm 2       $ & $ 19\pm 2       $ & 2002jul  \\
NGC6822     & S16        & $   3.7\pm 0.3 $ & $ 0.93\pm 0.09   $ & $  -76\pm 2       $ & $ 18\pm 2       $ & 2002jul  \\
NGC6822     & S30        & $   6.7\pm 0.2 $ & $ 1.32\pm 0.04   $ & $  -64.1\pm 0.7   $ & $ 27.7\pm 0.9   $ & 2005jul  \\
NGC6822     & S33        & $   7.0\pm 0.4 $ & $ 1.17\pm 0.08   $ & $  -44\pm 1       $ & $ 24\pm 2       $ & 2002jul  \\
Sagittarius & He 2-436   & $ 364\pm 1     $ & $ 0.820\pm 0.003 $ & $  129.16\pm 0.05 $ & $ 14.43\pm 0.06 $ & 2004juna \\
Sagittarius & StWr 2-21  & $ 171.4\pm 0.7 $ & $ 1.248\pm 0.005 $ & $  131.1\pm 0.1   $ & $ 25.9\pm 0.1   $ & 2004junc \\
Sagittarius & Wray 16-42 & $ 804\pm 4     $ & $ 1.057\pm 0.005 $ & $  128.8\pm 0.1   $ & $ 21.0\pm 0.1   $ & 2004juna \\
    \bottomrule
    \tabnotetext{a}{The fluxes are not calibrated.  See \S \ref{observations} and \S \ref{analysis}.}
    \tabnotetext{b}{The uncertainties listed are formal uncertainties.  See \S \ref{radial_velocities}.}
  \end{tabular}
\end{table*}

\begin{table*}[!t]\centering
  \setlength{\tabnotewidth}{0.7\columnwidth}
  \tablecols{7}
  \caption{[\ion{O}{3}]$\lambda$5007 Data for compact \ion{H}{2} regions} \label{tab_cHII}
 \begin{tabular}{llccccl}
    \toprule
Galaxy  & Object & Flux\ $^{\mathrm a}$   & FWHM  & $V_{helio}\ ^{\mathrm b}$ & $\Delta V_{0.5}$ & Run     \\
        &        & $(10^3\,\mathrm{ADU})$ & (\AA) & (km/s)      & (km/s)           &         \\
    \midrule
NGC6822 & S10    & $8.4\pm 0.4$ & $0.59\pm 0.03$ & $-88.1\pm 0.8$ & $16\pm 1    $ & 2002jul \\
NGC3109 & PN07   & $6.9\pm 0.2$ & $0.47\pm 0.02$ & $385.7\pm 0.4$ & $12.6\pm 0.5$ & 2004nov \\
NGC3109 & PN10   & $2.7\pm 0.2$ & $0.60\pm 0.06$ & $409\pm 2    $ & $17\pm 2    $ & 2004dec \\
    \bottomrule
    \tabnotetext{a}{The fluxes are not calibrated.  See \S \ref{observations} and \S \ref{analysis}.}
    \tabnotetext{b}{The uncertainties listed are formal uncertainties.  See \S \ref{radial_velocities}.}
  \end{tabular}
\end{table*}

\end{document}